\documentclass[letterpaper]{elsart5p}
\usepackage{amsmath,amssymb,graphicx}
\usepackage{lineno}
\usepackage{subfigure}
\usepackage[margin=1in]{geometry}

\newcommand{\ev}{\vec{e}}
\newcommand{\kv}{\vec{k}}
\newcommand{\m}{\vec{m}}
\newcommand{\x}{\vec{x}}
\newcommand{\Vm}{V_\textrm{max}}
\newcommand{\V}{\vec{V}}
\newcommand{\U}{\vec{U}}
\newcommand{\wl}{\omega_{\textrm{L}}}
\newcommand{\wnl}{\omega_{\textrm{NL}}}
\newcommand{\pd}[2]{\frac{\partial #1}{\partial #2}}

\newcommand{\eps}{\varepsilon}

\newcommand{\R}{\mathbb{R}}

\newcommand{\xiv}{\boldsymbol{\xi}}
\newcommand{\Pv}{\boldsymbol{\mathcal{P}}}
\newcommand{\Pvx}{\mathcal{P}_1}

\journal{}

\begin{document}
\date{}

\begin{frontmatter}
  \title{Propagating two-dimensional magnetic droplets}

  \author[NCSU]{M.~A.~Hoefer\ead{mahoefer@ncsu.edu}} and
  \author[NCSU]{M.~Sommacal\ead{msommac@ncsu.edu}}
  \address[NCSU]{Department of Mathematics, North Carolina State
    University, Raleigh, 27695-8205 NC, USA}

  \begin{abstract}
    Propagating, solitary magnetic wave solutions of the
    Landau-Lifshitz equation with uniaxial, easy-axis anisotropy in
    thin (two-dimensional) magnetic films are investigated.  These
    localized, nontopological wave structures, parametrized by their
    precessional frequency and propagation speed, extend the
    stationary, coherently precessing ``magnon droplet'' to the moving
    frame, a non-trivial generalization due to the lack of Galilean
    invariance.  Propagating droplets move on a spin wave background
    with a nonlinear droplet dispersion relation that yields a limited
    range of allowable droplet speeds and frequencies.  An iterative
    numerical technique is used to compute the propagating droplet's
    structure and properties.  The results agree with previous
    asymptotic calculations in the weakly nonlinear regime.
    Furthermore, an analytical criterion for the droplet's orbital
    stability is confirmed.  Time-dependent numerical simulations
    further verify the propagating droplet's robustness to
    perturbation when its frequency and speed lie within the allowable
    range.
  \end{abstract}

\end{frontmatter}

\section{Introduction\label{sec:introduction}}

Magnetic materials yield a rich variety of intriguing nonlinear wave
phenomena.  Recent theoretical and experimental developments
\cite{slonczewski_current-driven_1996,tsoi_excitation_1998,rippard_direct-current_2004,silva_developments_2008}
have enabled the controlled manipulation of magnetic moments on the
nanometer length scale, the magnetic exchange length, thereby
generating further interest in the field of nanomagnetism
\cite{lau_magnetic_2011}.  Dynamic, strongly nonlinear, localized wave
structures varying on the exchange length scale have been studied
theoretically for some time \cite{ivanov_bound-states_1977} (see the
exhaustive review \cite{kosevich_magnetic_1990}).  It was recently
proposed that one member of this family of solitary waves, the
two-dimensional (2D) nontopological droplet, could be experimentally
realized in a spin torque nanocontact system that delivers a localized
torque to a ferromagnetic thin film, balancing the inherent material
damping \cite{hoefer_theory_2010}.  The so-called dissipative droplet
can, under certain conditions, experience a drift instability,
propagating for some time before eventually succumbing to magnetic
damping.  A natural question arises, then, to inquire into the
existence, stability, and properties of propagating, nontopological
solitary waves in the underlying conservative model, the
Landau-Lifshitz (LL) equation.

Previous studies of 2D propagating droplets include numerical
computations in the isotropic \cite{cooper_solitary_1998} and
anisotropic \cite{piette_localized_1998} regimes as well as weakly
nonlinear asymptotics \cite{ivanov_zaspel_yastremsky_2001}.  In
\cite{piette_localized_1998}, a numerical study of topological and
nontopological localized solutions in 2D ferromagnets with easy-axis
anisotropy was undertaken.  Stationary droplets were constructed
numerically which were then made to propagate in time-dependent
simulations by the imposition of an initial, unidirectional phase
gradient.  The following approximate properties were observed: the
excitation, accompanied by spin wave radiation, moved coherently with
constant speed, a single frequency of precession, and a spatially
dependent phase lag.  In this work, we demonstrate that the
excitations observed in \cite{piette_localized_1998} and the moving,
coherent structures in \cite{hoefer_theory_2010} can be identified as
propagating, nontopological droplet solitary waves which we construct
directly by studying a nonlinear eigenvalue problem.

We undertake a careful numerical and asymptotic study of propagating
2D nontopological droplets, from here on in termed droplets.  A
two-parameter family of localized solitary wave solutions to the LL
equation with easy-axis anisotropy parametrized by speed $V$ and rest
frequency $\omega$ is constructed.  In order to avoid a linear
resonance, we are led to postulate the droplet's existence and
stability for speeds and frequencies in the restricted range
\begin{equation*}
  1 - \omega - \frac{V^2}{4} > 0,
\end{equation*}
similar to the existence conditions for moving 1D droplets
\cite{kosevich_magnetic_1990,long_nonlinear_1979}.  The droplet's
structure is studied analytically in the large amplitude, small speed
regime and in the weakly nonlinear regime following
\cite{ivanov_zaspel_yastremsky_2001}.  A nonlinear dispersion relation
for a background spin wave which modulates the propagating droplet
naturally arises.  It degenerates to the linear dispersion relation
for exchange spin waves in the small amplitude regime where the
Nonlinear Schr\"{o}dinger equation approximately governs the dynamics.
An iterative numerical method is employed to compute the droplet's
structure in the comoving frame by solving an appropriate boundary
value problem.  These results are used to compute the amplitude,
asymmetry, energy, momentum, and more for a wide range of propagating
droplets which agree with the asymptotic calculations.  Droplet
orbital stability is confirmed by numerical verification of an
analytical, Jacobian condition.  Time-dependent numerical simulations
with initially perturbed propagating droplets further demonstrate the
robustness of droplet propagation.


The outline of this work is as follows.
In Sec.~\ref{sec:modelequations}, several equivalent forms of the LL
equation are introduced for later use in asymptotic and numerical
analysis.  In addition, properties of the stationary droplet are
reviewed.  The propagating droplet problem formulation and some basic
properties are discussed in Sec.~\ref{sec:movingdroplet}. Section
\ref{sec:prop-dropl-asympt} presents the asymptotic results.
In Sec.~\ref{sec:moving-dropl-numer}, numerical results addressing
droplet properties and stability are given.  Further discussion and
conclusions are related in Sec.~\ref{sec:discussion} and
\ref{sec:conclusions}.  Finally, validation of the numerical technique
is presented in the Appendix.

\section{Model Equations and the Stationary
  Droplet\label{sec:modelequations}}

The Landau-Lifshitz torque equation
\begin{subequations}
  \label{eq:54}
  \begin{align}
    &\frac{\partial \vec{M}}{\partial t} = - | \gamma | \mu_0 \vec{M}
    \times \vec{H}_{\textrm{eff}} \, , \\
    \vec{H}_{\textrm{eff}} &= \\
    \nonumber
    & ~\frac{2 A }{\mu_0 M_{\mathrm{s}}^2} 
    \Delta \vec{M} + \left ( H_0 + \frac{2K_{\mathrm{u}}}{\mu_0
        M_{\mathrm{s}}^2} M_3 
    \right ) \vec{e}_3 + \vec{H}_d \, ,
  \end{align}
\end{subequations}
describes the dissipationless dynamics of the magnetization
$\vec{M}: ~\R^2 \times [-\delta/2,\delta/2] \times \R^+ \to
M_{\mathrm{s}} \mathbb{S}^2$ in an unbounded ferromagnetic film of
thickness $\delta$ exhibiting perpendicular, uniaxial anisotropy and
a perpendicular applied magnetic field \cite{landau_theory_1935}.

Relevant parameters for Eq.~(\ref{eq:54}) are the gyromagnetic ratio
($\gamma$), the free space permeability ($\mu_0$), the exchange
stiffness parameter ($A$), the perpendicular magnetic field amplitude
($H_0$), and the crystalline anisotropy constant ($K_{\mathrm{u}}$).

\begin{figure}
  \centering
  \includegraphics[width=0.99\columnwidth]{
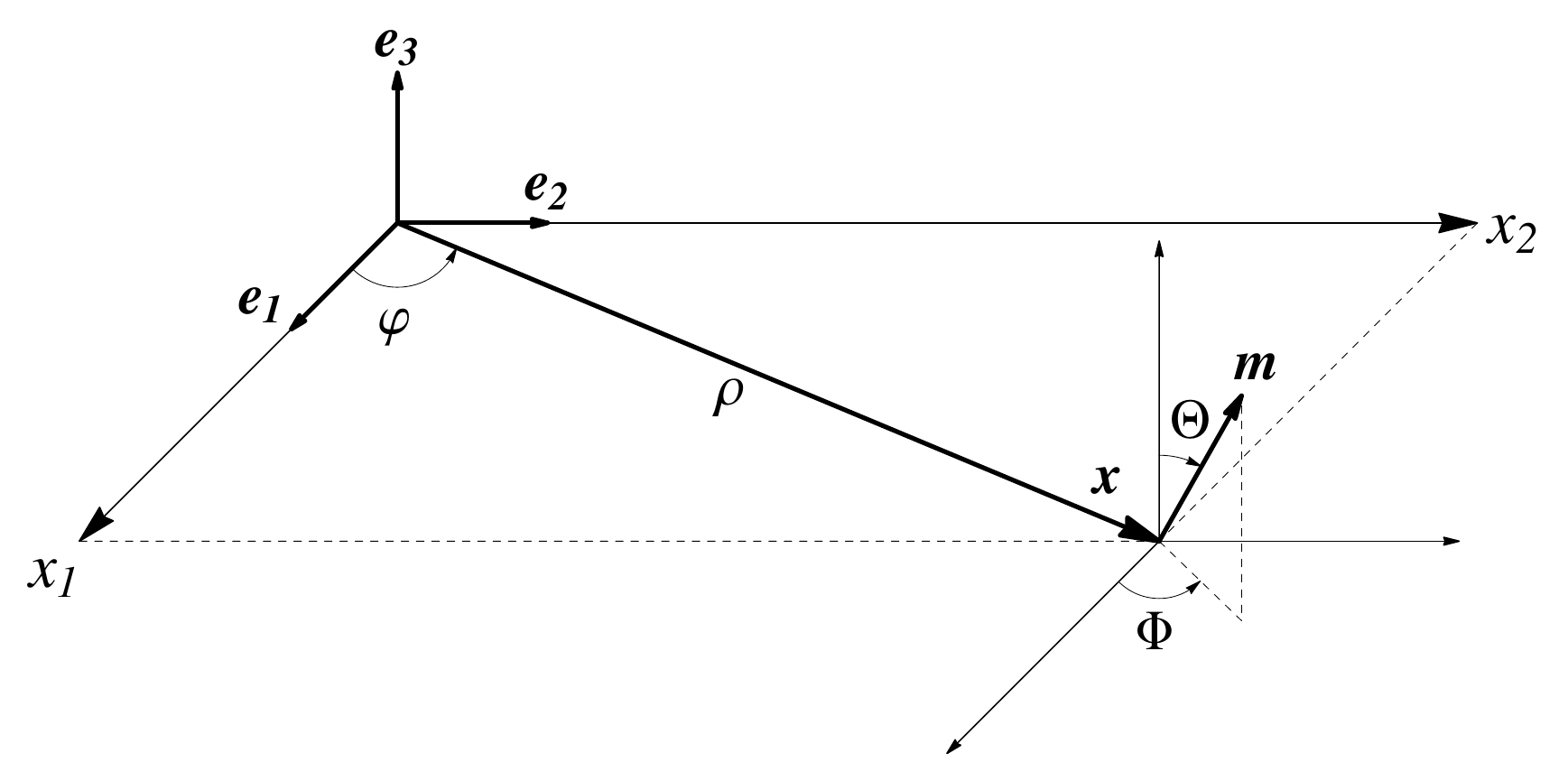}
  \caption{Schematic of the coordinate systems for a planar
    ferromagnet consisting of the domain of $\vec{m}$
    ($\mathbb{R}^{2}$ in polar coordinates with radius $\rho$,
    azimuthal angle $\varphi$) and the range of $\vec{m}$ (unit sphere
    $\mathbb{S}^{2}$ with polar angle $\Theta$, azimuthal angle
    $\Phi$).  \label{fig:refsys}}
\end{figure}

The magnetization
\begin{equation}
  \label{eq:58}
  \vec{M} \equiv M_{\mathrm{s}}
  \m \equiv M_{\mathrm{s}} \sum_{j=1}^{3}m_{j}\,\vec{e}_{j}\, ,
\end{equation}
satisfies $|\vec{M}(\vec{x},t)| = M_{\mathrm{s}}$, where
$M_{\mathrm{s}}$ is the saturation magnetization, as can be verified
by taking the dot product of (\ref{eq:54}) with $\vec{M}$.  The
vectors $\vec{e}_j$, $j =1,2,3$, are the standard Cartesian basis
vectors for $\R^3$ (see Fig.~\ref{fig:refsys}).  The magnetostatic or
dipolar field $\vec{H}_d$ is, in general, nonlocal, coupling the
magnetization dynamics to Maxwell's equations which, in the
magnetostatic regime, take the form
\begin{equation*}
  \nabla \cdot \vec{H}_d = - \nabla \cdot \vec{M},
\end{equation*}
where $\vec{M}$ is extended to zero outside of $\R^2 \times
[-\delta/2,\delta/2]$.  We are interested in the thin film regime
where $\delta$ is sufficiently small.  There are a number of different
thin film scalings that depend upon the length scales inherent to the
problem at hand.  In the static case, some thin film limits have been
rigorously justified via $\Gamma$-convergence; see, e.g.,
\cite{kohn_effective_2005,desimone_recent_2006} and references
therein.  Thin film magnetodynamics have also been rigorously studied
\cite{kohn_effective_2005,melcher_thin-film_2010}.  In many thin film
scalings, there are two key simplifications that can be made:
the magnetization is approximately uniform through the thickness
of the film so that the problem is approximately 2D
\begin{equation}
  \label{eq:57}
  \Delta \to \frac{\partial^2}{\partial x_1^2} +
  \frac{\partial^2}{\partial x_2^2} ,
\end{equation} and the dipolar field  can be approximated by the local term
\begin{equation}
  \label{eq:56}
  \vec{H}_d \to -M_3 \vec{e}_3 .
\end{equation}
Heuristically, the two-dimensionality of the problem is justified when
the film thickness is much smaller than the transverse magnetic
excitations of interest because the exchange energy for magnetization
variations in the $x_3$ direction would be prohibitively large.  In
our case, this amounts to the requirement
\begin{equation}
  \label{eq:71}
  \delta \ll L_{ex}/(Q - 1)^{1/2}\, ,
\end{equation}
where $L_{ex}$ is the magnetic exchange length
\begin{equation*}
  L_{ex} = \left (\frac{2A}{\mu_0 M_{\mathrm{s}}^2}
  \right )^{1/2}\, ,
\end{equation*}
typically on the order of several nanometers, and $Q$ is the
dimensionless quality factor
\begin{equation*}
  Q = \frac{2 K_{\mathrm{u}}}{\mu_0
    M_{\mathrm{s}}^2} \, ,
\end{equation*}
assumed here to be greater than unity.  Assuming that $\partial_{x_3}
\vec{M} = 0$, we can compute the 2D Fourier transform of the $x_3$
averaged dipolar field (see
e.g.~\cite{garcia-cervera_one-dimensional_2004})
\begin{equation}
  \label{eq:79}
  \begin{split}
    \widehat{(\vec{H}_d + M_{\mathrm{s}} \vec{e}_3)} = &~ -
    \frac{\vec{k} [ \vec{k} \cdot \widehat{\vec{M}}_\perp(\vec{k})
      ]}{k^2} [1 - \widehat{\Gamma}(k \delta)] \\
    &~- \widehat{(M_3 - M_{\mathrm{s}})}(\vec{k})\, \widehat{\Gamma}(k
    \delta) \vec{e}_3 \, ,
  \end{split}
\end{equation}
where $\widehat{f}(\vec{k}) = \int_{\R^2} f(\vec{x}_\perp) e^{-i
  \vec{k} \cdot \vec{x}_\perp} \, d\vec{x}_\perp$, $\vec{x}_\perp =
(x_1,x_2)$, $\vec{M}_\perp = (M_1,M_2)$, and
\begin{equation}
  \label{eq:76}
  \widehat{\Gamma}(k\delta) = \frac{1 - e^{-k \delta}}{k \delta}\,
  . 
\end{equation}
The Fourier transforms are computed for rapidly decaying functions.
In this work, we study solutions that rapidly decay to $\vec{M} \to
M_{\mathrm{s}} \vec{e}_3$ with the decay length $L_{ex}/(Q-1)^{1/2}$
(see non-dimensionalization below).  Then the Fourier transforms in
\eqref{eq:79} are concentrated on wavevectors $\vec{k}$ of order
$(Q-1)^{1/2}/L_{ex}$ or less.  Under the assumption \eqref{eq:71}, we
can approximate \eqref{eq:79}, \eqref{eq:76} as
\begin{equation*}
  \begin{split}
    \widehat{\Gamma}(k\delta) &\sim 1 - \frac{1}{2} k \delta = 1 +
    \mathcal{O}\left ( \delta \frac{(Q-1)^{1/2}}{L_{ex}} \right ),
    \\
    \vec{H}_d &= -M_3 \vec{e}_3 + \mathcal{O} \left ( \delta
      \frac{(Q-1)^{1/2}}{L_{ex}} \right ),
  \end{split}
\end{equation*}
hence \eqref{eq:56} is justified.

It is convenient to nondimensionalize Eq.~(\ref{eq:54}) according to
(\ref{eq:58}) and
\begin{equation}
  \label{eq:59}
  \begin{split}
    t' &= | \gamma | \mu_0 M_{\mathrm{s}} (Q - 1) t \, , \quad
    \vec{x}' = \frac{(Q - 1)^{1/2}}{L_{ex}} \vec{x} \, , \\
    h_0 &= \frac{H_0}{M_{\mathrm{s}}(Q - 1)} ,
  \end{split}
\end{equation}
so that, after dropping primes, the problem of studying localized
solutions to Eq.~(\ref{eq:54}) subject to (\ref{eq:57}) and
(\ref{eq:56}) becomes
\begin{equation}
\label{eq:landaulifshitz-vector-withparameters}
\begin{split}
  & \frac{\partial \vec{m}}{\partial t} = - \vec{m} \times
  \left[\Delta \vec{m} + (m_3+h_{0})\, \vec{e}_3\right]\,\\%
  & \vec{m}: ~ \mathbb{R}^2 \times \mathbb{R}^+ \to \mathbb{S}^2,
  \quad%
  \lim_{|\vec{x}| \to \infty} \vec{m}(\vec{x},t) = \vec{e}_3, \quad t
  \geq 0 .
\end{split}
\end{equation}
The nondimensionalization (\ref{eq:59}) assumes an easy-axis
anisotropy so that the quality factor $Q$, measuring the relative
strength of the anisotropy to the magnetostatic energy, is larger than
unity.  When $Q < 1$, the anisotropy is of the easy-plane variety.
Anisotropic ferromagnetic nanostructures in this thin film geometry
can be fabricated, e.g., using Cobalt/Nickel multilayers
\cite{lau_magnetic_2011}.

Transforming to the rotating frame
\begin{equation*}
  \begin{split}
    \m' = [&m_1\,\cos{(h_0\,t)} + m_2\,\sin{(h_0\,t)}, \\
    &- m_1 \sin{(h_0\,t)} + m_2\,\cos{(h_0\,t)}, m_3] ,
  \end{split}
\end{equation*}
takes Eq.~(\ref{eq:landaulifshitz-vector-withparameters}) to the
parameterless form (after dropping the prime)
\begin{equation}
  \label{eq:landaulifshitz-vector}
    \frac{\partial \vec{m}}{\partial t} = - \vec{m} \times
    \vec{h}_{\textrm{eff}}, \quad \vec{h}_{\textrm{eff}} = \Delta
    \vec{m} + m_3\,\vec{e}_3  .
\end{equation}
Therefore, without loss of generality, we will consider localized
solutions of Eq.~(\ref{eq:landaulifshitz-vector}).

Equation \eqref{eq:landaulifshitz-vector} admits a family of
localized, coherently precessing solutions
\cite{ivanov_bound-states_1977,kosevich_magnetic_1990}.  The solutions
that incorporate a complete reversal of the magnetization,
i.e.~$\m(\x_0,t) = \mp \ev_3$ and $\lim_{|\x| \to \infty} \m(\x,t) =
\pm \ev_3$ for one $\x_0 \in \R^2$ (magnetic vortices), are called
topological and exhibit pinning of the soliton center, hence cannot
propagate \cite{papanicolaou_dynamics_1991}.  In contrast, solutions
that do not exhibit magnetization reversal are termed nontopological
and can propagate.  Such solutions are the focus of this work.

Sufficiently localized solutions of
Eq.~(\ref{eq:landaulifshitz-vector}) conserve the magnetic energy
\begin{equation}
  \label{eq:3}
  {\cal E}[\vec{m}] = \frac{1}{2} \int \left [
  \left|\nabla \vec{m}\right|^2 + \left(1-m_3^2\right) \right ]\,
  \mathrm{d}\vec{x}\,,
\end{equation}
which incorporates contributions due to exchange $|\nabla \m|^2$ and
anisotropy $(1-m_3^2)$.  Integration is always assumed to be taken
over the plane $\R^2$ unless otherwise noted.  The effective field in
Eq.~(\ref{eq:landaulifshitz-vector}) is derived from the magnetic
energy via
\begin{equation*}
  \vec{h}_{\textrm{eff}} = - \frac{\delta \mathcal{E}}{\delta \m} \, .
\end{equation*}

Other conserved quantities include the spin density
\begin{equation}
  \label{eq:4}
  \mathcal{N}[\m] = \int (1 - m_3) \, \mathrm{d}\x,
\end{equation}
and the momentum
\begin{equation}
  \label{eq:5}
  \Pv[\m] = \int \frac{m_2 \nabla m_1 - m_1 \nabla m_2}{1 + m_3} \,
  \mathrm{d}\x . 
\end{equation}
The quantity ${\cal N}$ can be interpreted as the mean number of spin
deviations from the uniform state $\ev_3$ in a localized magnetic
excitation.  In the quantum limit it takes integer values and
corresponds to the number of magnons in an excited ferromagnet
\cite{kosevich_magnetic_1990}.

Note that the momentum as given in \eqref{eq:5} is ambiguous when the
magnetization involves topological structures.  The ambiguity has been
addressed by considering moments of the topological vorticity
\cite{papanicolaou_dynamics_1991}.  Since we are considering
nontopological structures, the momentum in \eqref{eq:5} is equivalent
to the more general formulation of \cite{papanicolaou_dynamics_1991}.

Small amplitude exchange spin waves deviating from the background
state $\ev_3$ in the form
\begin{equation*}
  \begin{split}
    m_1 &= \varepsilon \cos (\kv \cdot \x - \wl t),\\
    m_2 &= - \eps \sin (\kv \cdot \x - \wl t)], \quad
    m_3 \sim 1, \quad
    | \varepsilon |\ll 1\,,
  \end{split}
\end{equation*}
with wavenumber $\kv$ and frequency $\wl$ satisfy the linear
dispersion relation
\begin{equation}
  \label{eq:dispersionrelation-linear}
  \wl(\vec{k})\equiv\wl (k)=1+k^{2}\,,\quad k=|\vec{k}|\,.
\end{equation}


\subsection{Alternate Forms of the Landau-Lifshitz Equation}
\label{sec:altern-forms-land}

Two alternative forms for Eq.~\eqref{eq:landaulifshitz-vector} will be
used in this work.  For the first, we use the spherical basis (see
Fig.~\ref{fig:refsys})
\begin{equation*}
  \m = [\cos{\Phi}\,\sin{\Theta}, \sin{\Phi}\,\sin{\Theta}, \cos{\Theta}],
\end{equation*}
where $\Phi \in [0,2\pi)$ $(\textrm{mod}\, 2\pi)$ and $\Theta \in
[0,\pi]$ are the azimuthal and polar angles, respectively.
Inserting this expression into Eq.~(\ref{eq:landaulifshitz-vector})
gives
\begin{subequations}
  \label{eq:landaulifshitz-spherical}
  \begin{equation}
    \label{eq:landaulifshitz-spherical-a} \frac{\partial
      \Theta}{\partial t} =
    F[\Theta,\Phi]\,,\qquad \sin{\Theta} \frac{\partial \Phi}{\partial
      t} = G[\Theta,\Phi]\,,%
  \end{equation}
  where
  \begin{equation}
    \label{eq:landaulifshitz-spherical-b}
    \begin{split}
      F[\Theta,\Phi] &= \frac{\nabla \cdot \left ( \sin^2{\Theta}\, \nabla
          \Phi \right )}{\sin{\Theta}},\\
      G[\Theta,\Phi] &= \frac{1}{2}\,\sin(2\Theta)\, ( |\nabla
      \Phi|^2 + 1 )
      - \Delta \Theta\,.
    \end{split}
  \end{equation}
\end{subequations}
We will see that Eq.~(\ref{eq:landaulifshitz-spherical}) is in a
useful form to study stationary and slowly moving droplets as well as
to formulate an energy minimization problem for propagating droplets.
The energy, spin density, and momentum in these variables take the
form
\begin{subequations}
  \label{eq:16}
  \begin{align}
    \label{eq:11}
    {\cal E}[\Theta,\Phi] &= \frac{1}{2} \int \left [ \left|\nabla
        \Theta\right|^2 + \sin^2{\Theta}\, \left(1+\left|\nabla
          \Phi\right|^2\right) \right ] \, \mathrm{d}\vec{x}\,,
    \\
    \label{eq:12}
    \mathcal{N}[\Theta] &= \int \left (1 - \cos{\Theta} \right) \, \mathrm{d}\x\,, \\
    \label{eq:13}
    \Pv[\Theta,\Phi]&= \int \left ( \cos{\Theta} - 1 \right)
    \,\nabla{\Phi}\,\mathrm{d}\x\, .
  \end{align}
\end{subequations}
Note that \eqref{eq:landaulifshitz-spherical} is Hamilton's
equations in the form
\begin{equation*}
  \pd{\Phi}{t}  = -\frac{\delta \mathcal{E}}{\delta \cos{\Theta}}, \quad
  \pd{\cos{\Theta}}{t}  = \frac{\delta \mathcal{E}}{\delta \Phi},
\end{equation*}
where $\Phi$ and $\cos{\Theta}$ are the canonically conjugate
variables for the Hamiltonian system.


Another representation of (\ref{eq:landaulifshitz-vector}) can be
had by performing the stereographic projection
\begin{equation}
  \label{eq:eq:stereographicprojection}
  w = \frac{m_1 + i m_2}{1 + m_3} = \frac{e^{i \Phi}\sin{\Theta}}{1 +
    \cos{\Theta}}.
\end{equation}
Then \eqref{eq:landaulifshitz-vector} becomes
\cite{lakshmanan_landau-lifshitz_1984}
\begin{equation}
  \label{eq:landaulifshitz-stereographic}
  i\,w_t = \Delta w - \frac{2 w^* \nabla w \cdot \nabla w + w
    (1 - |w|^2)}{1 + |w|^2}\,,
\end{equation}
where the appended asterisk denotes complex conjugation.  This form of
the LL equation is amenable to weakly nonlinear asymptotics
(Sec.~\ref{sec:prop-dropl-asympt}) and numerical analysis
(Sec.~\ref{sec:moving-dropl-numer}).  The conserved quantities are
\begin{subequations}
  \label{eq:49}
  \begin{align}
    \label{eq:14}
    {\cal E}[w] &= 2 \int \frac{\left|\nabla w \right|^2 +
      |w|^2}{\left(1 + |w|^2\right)^2} \, \mathrm{d} \vec{x}\, ,\\
    \label{eq:17}
    \mathcal{N}[w] &= 2 \int \frac{|w|^2}{1 + |w|^2} \, \mathrm{d}\x\, , \\
    \label{eq:18}
    \Pv[w] &= -2 \int \frac{\textrm{Im}(w^* \nabla w)}{1 + |w|^2}
    \, \mathrm{d}\vec{x}\,.
  \end{align}
\end{subequations}
In analogy with a Bose-Einstein condensate
\cite{pethick_bose-einstein_2002}, the momentum (\ref{eq:18})
corresponds to a weighted phase gradient or a ``magnetic superfluid''
momentum.

For the time-dependent initial value problem discussed in Sec.
\ref{sec:droplet-stability}, we will solve Eq.
\eqref{eq:landaulifshitz-vector}.  Thus, we will make use of all
three forms \eqref{eq:landaulifshitz-vector},
\eqref{eq:landaulifshitz-spherical}, and
\eqref{eq:eq:stereographicprojection} of the LL equation in our
analysis.

\subsection{Stationary Droplet\label{sec:stationarydroplet}}%

An azimuthally symmetric, stationary, localized droplet solution to
the LL equation was studied in
\cite{ivanov_bound-states_1977,kosevich_magnetic_1990}.  We note that
the term magnon droplet was used because of its analogy with a
localized ``droplet'' of a condensed, attractive one-dimensional Bose
gas \cite{kosevich_magnetic_1990}.  In this section, we briefly review
the properties of the stationary droplet.

It is convenient to write the stationary droplet solution in a
spherical basis satisfying Eq.~\eqref{eq:landaulifshitz-spherical}.
It takes the form
\begin{equation*}
  \begin{split}
    &\Theta(\vec{x},t)=\Theta_0(\rho;\omega)\,,\quad\Phi(\vec{x},t)=\omega\,t\,
    ,\\
    &F[\Theta_0,\omega\,t]=0\,,\quad G[\Theta_0,\omega t]=\omega\,\sin{\Theta_0}\,,
   \end{split}
\end{equation*}
where $\rho = \sqrt{x_1^2 + x_2^2}$ is the distance from the origin
and the polar angle function $\Theta_0$, parametrized by the
droplet frequency $\omega$, is determined by solving the following
nonlinear eigenvalue problem
\begin{subequations}
\label{eq:Theta0-equation}
\begin{flalign}
  \label{eq:43}
  &-\left ( \frac{\mathrm{d}^2}{\mathrm{d} \rho^2} + \frac{1}{\rho}
    \frac{\mathrm{d}}{\mathrm{d} \rho} 
  \right ) \Theta_0 + \sin{\Theta_0}\,\cos{\Theta_{0}} - \omega
  \sin{\Theta_0} = 0\, ,\\
  &\frac{\mathrm{d} \Theta_0}{\mathrm{d} \rho}(0;\omega) = 0 \,, \quad
  \lim_{\rho \to 
    \infty} \Theta_0(\rho; \omega) = 0 \,.
\end{flalign}
\end{subequations}
The ground state solution that is positive and monotonically decaying
is sought. One can show that $0 < \Theta_0(\rho;\omega) < \pi$,
therefore this solution is termed \textit{non-topological}
\cite{kosevich_magnetic_1990}.  In contrast, there exist
\textit{topological} droplets that satisfy a related nonlinear
eigenvalue problem with the conditions $\Theta_0(0;\omega) = \pi$,
$\Theta_0(\rho;\omega) \to 0$.  Their existence and stability subject
to certain symmetric perturbations were studied rigorously in
\cite{gustafson_stability_2002}.

Assuming the existence of a localized solution, the following
inequality holds
\begin{equation}
  \label{eq:omega-interval}
  0 < \omega < 1 \, ,
\end{equation}
which can be seen by multiplying Eq.~(\ref{eq:43}) by
$\frac{\mathrm{d} \Theta_0}{\mathrm{d} \rho}\,\rho^2$ and
integrating to find
\begin{equation}
  \label{eq:Theta0-integralidentity}
  2\,\omega
  \int_0^\infty \left(1 - \cos{\Theta_0}\right)\, \rho \, \mathrm{d}\rho -
  \int_0^\infty \sin^2{\Theta_0}\,\, \rho \, \mathrm{d}\rho = 0 .
\end{equation}
We immediately observe that
\begin{equation*}
  \omega =  \frac{ \int_0^\infty \sin^2{\Theta_0}\,\, \rho
    \, \mathrm{d}\rho}{2 \int_0^\infty (1 - \cos{\Theta_0})\, \rho \, \mathrm{d}\rho}
  > 0\,.
\end{equation*}
Using the fact that $2\,(1-\cos{\Theta}) > \sin^2{\Theta}$ for $0 <
\Theta < \pi$, we also have $\omega < 1$.


The decay rate of $\Theta_{0}$ can be obtained by expanding
(\ref{eq:43}) for large values of $\rho$ and neglecting
nonlinear terms which gives
\begin{equation*}
  -\frac{\mathrm{d}^{2}
\Theta_0}{\mathrm{d}\rho^{2}}-\frac{1}{\rho} \frac{\mathrm{d}
\Theta_0}{\mathrm{d}\rho}+(1-\omega)\,\Theta_{0} \sim 0\,,
\end{equation*}
giving
\begin{equation}
  \label{eq:68}
  \Theta_{0}(\rho) =
  \mathcal{O}\left(\frac{e^{-\rho\,\sqrt{1-\omega}}}{\sqrt{\rho}}\right)\,,\quad\rho\gg
  1\,.
\end{equation}

From numerical computations, see, e.g., \cite{piette_localized_1998},
it is known that the amplitude of the droplet increases with
decreasing frequency.  As the frequency approaches zero, the droplet
resembles a circular domain wall with a transition layer that moves
further away from the droplet center.  The nonlinear, orbital
stability of nontopological, three-dimensional solitary waves in the
context of anisotropic, two sublattice ferrites were studied by use of
a Lyapunov functional whose positive definiteness is guaranteed when
\cite{ivanov_three-dimensional_1984}
\begin{equation}
  \label{eq:87}
  \frac{\d \mathcal{N}}{\d \omega} < 0\, ,
\end{equation}
which is commonly referred to as a slope or VK condition after
Vakhitov-Kolokolov's work on Nonlinear Schr\"{o}dinger solitary waves
\cite{Vakhitov-73}.  Since $\mathcal{N}'(\omega)$ is the Jacobian of
the map from the physical variable $\omega$ to the conserved quantity
$\mathcal{N}$, we refer to (\ref{eq:87}) as a Jacobian condition as
this will naturally generalize to the stability criterion for
propagating droplets.  The Jacobian condition (\ref{eq:87}) has been
numerically verified for 2D stationary droplets
\cite{kosevich_magnetic_1990}, suggesting their orbital stability.
Their robustness to perturbations in 2D, time-dependent numerical
simulations \cite{hoefer_theory_2010} are further evidence of their
stability.

\section{Propagating Droplet: Problem Formulation and Basic
  Results \label{sec:movingdroplet}}%

A two-parameter family of moving droplet solutions of
\eqref{eq:landaulifshitz-spherical} is considered in the form
\begin{subequations}
\label{eq:8}
\begin{align}
&\Theta(\x,t) = \theta(\xiv;\omega,\V)\,,\\
&\Phi(\x,t) = \omega t + \phi(\xiv;\omega,\V)\,,
\end{align}
with
\begin{equation}
\xiv = \x - \V\,t \, ,
\end{equation}
\end{subequations}
where $\omega$ is the droplet's rest frequency and $\V$ is the
droplet's velocity.  Insertion of this ansatz into
Eq.~\eqref{eq:landaulifshitz-spherical} results in
\begin{subequations}
  \label{eq:23}
  \begin{multline}
    \label{eq:20}
    \sin{\theta}\,\left ( \omega - \V \cdot \nabla \phi \right ) =\\ -
    \Delta \theta + \sin{\theta}\,\cos{\theta}\,
    \left ( 1 + | \nabla \phi |^2 \right )\,,
  \end{multline}
  \begin{align}
    \label{eq:24}
    &- \sin{\theta}\,\V \cdot \nabla \theta = \nabla \cdot \left (\sin^2
      \theta\,\nabla \phi \right ) .
  \end{align}
\end{subequations}
The propagating nontopological droplet that we seek satisfies the
boundary conditions
\begin{equation}
  \label{eq:22}
  \lim_{|\xiv| \to \infty} \theta(\xiv;\omega,\V) = 0\,,\,\,
  \lim_{|\xiv| \to \infty} \nabla \phi(\xiv;\omega,\V) = \textrm{const}\,,
\end{equation}
with $\theta(\xiv;\omega,\V) < \pi$ for all $\xiv \in \R^2$.
Assuming decay for $|\xiv|$ sufficiently large, $\theta \ll 1$ and
$\nabla \phi \sim \U$ a constant vector, Eq.~\eqref{eq:24} is
asymptotically
\begin{equation*}
  -\V \cdot \nabla \theta = 2 \U \cdot \nabla \theta\, .
\end{equation*}
Therefore,
\begin{equation}
  \label{eq:26}
  \lim_{|\xiv| \to \infty} \nabla \phi(\xiv;\omega,\V) = -\frac{\V}{2}
  \, .
\end{equation}

Due to the $SO(2)$ invariance of the LL equation
\eqref{eq:landaulifshitz-vector}, it is sufficient to consider
\begin{equation}
  \label{eq:30}
  \V = V \ev_1 , \quad V > 0 \, .
\end{equation}
For the rest of this work, expression \eqref{eq:30} is assumed to
hold.

By use of (\ref{eq:16}), we observe that (\ref{eq:23}) is equivalent
to the following
\begin{subequations}
  \label{eq:21}
  \begin{align}
      \frac{\delta}{\delta \theta} \left ( \omega\, \mathcal{N} + V\,
        \Pvx \right ) &= \frac{\delta \mathcal{E}}{\delta \theta}\, , \\
      \frac{\delta}{\delta \phi} \left ( \omega\, \mathcal{N} + V\,
        \Pvx \right ) &= \frac{\delta \mathcal{E}}{\delta \phi} \, ,
  \end{align}
\end{subequations}
where $\mathcal{P}_{1}$ is the nonvanishing component of
$\vec{\mathcal{P}}$ when condition (\ref{eq:30}) holds. Equations
(\ref{eq:21}) are the necessary conditions for a minimizer of the
functional
\begin{equation}
  \label{eq:27}
  \mathcal{D}[\theta,\phi] = \mathcal{E}[\theta,\phi] - \omega
  \mathcal{N}[\theta,\phi] - V
  \Pvx[\theta,\phi] \, ,
\end{equation}
where $\omega$ and $V$ are Lagrange multipliers ensuring the
constraints of fixed spin density $\mathcal{N}$ and momentum $\Pvx$.
Propagating, localized structures in 2D isotropic
\cite{cooper_solitary_1998} and 3D anisotropic
\cite{ioannidou_soliton_2001} ferromagnets were studied numerically by
minimizing a similar functional.  Note that,
\begin{equation}
  \label{eq:37}
  \mathcal{N} = -\frac{\partial \mathcal{D}}{\partial \omega}\, , \quad
  \mathcal{P}_1 = -\frac{\partial \mathcal{D}}{\partial V}\, .
\end{equation}
While the variational formulation of the droplet leads to a
parameterization in terms of $\mathcal{N}$ and $\mathcal{P}_1$, for
physical applications, it is natural to parameterize the droplet in
terms of the physical variables $\omega$ and $V$.  Solving
Eqs.~(\ref{eq:23}) for fixed $\omega$ and $V$ provides the map
$(\omega,V) \mapsto (\mathcal{N},\mathcal{P}_1)$.  Invertibility of
this map requires that the Jacobian
\begin{equation}
  \label{eq:83}
  J = \frac{\partial
    \mathcal{P}_1}{\partial V} \frac{\partial \mathcal{N}}{\partial
    \omega} - \frac{\partial \mathcal{N}}{\partial V} \frac{\partial
    \mathcal{P}_1}{\partial \omega} \,, 
\end{equation}
be nonzero.  In the context of other magnetic and nonlinear wave
models, the orbital stability of two-parameter solitary waves is
determined by the Jacobian condition
\cite{baryakhtar_vector_1983,buryak_stability_1996}
\begin{equation}
  \label{eq:88}
  J < 0\, .
\end{equation}
Note that the Jacobian condition (\ref{eq:88}) degenerates to the
Jacobian condition (\ref{eq:87}) for stationary droplets when $V = 0$
because $\partial_\omega \mathcal{P}_1 = 0$ and $\partial_V
\mathcal{P}_1 > 0$.  Droplet stability will be discussed further in
Secs.~\ref{sec:small-amplitude} and \ref{sec:droplet-stability}.

If $\theta_*(\xiv)$, $\phi_*(\xiv)$ are a minimizer of \eqref{eq:27},
then dilation of $\xiv$ by the factor $\lambda$ gives
\begin{equation*}
  d(\lambda) = \mathcal{D}[\theta_*(\lambda \cdot),\phi_*(\lambda
  \cdot)] \,.
\end{equation*}
A necessary condition for $\theta_*(\xiv)$, $\phi_*(\xiv)$ to be a
minimum of \eqref{eq:27} is therefore
\begin{align}
  \label{eq:29}
  d'(1) &= \int \sin^2{\theta_*} \,\,\mathrm{d}\xiv - V\, \Pvx - 2\,
  \omega\, \mathcal{N} = 0 \, ,
\end{align}
which can be derived more generally by direct manipulation of
conservation laws for the Landau-Lifshitz equation
\cite{papanicolaou_semitopological_1999}.  Similar identities were
studied by Derrick \cite{derrick_comments_1964} and Pohozaev
\cite{pohozaev_eigenfunctions_1965}.  We will make use of this
relation in order to verify the validity of our numerical computations
in the Appendix.

Another formulation of the boundary value problem \eqref{eq:23},
\eqref{eq:22}, (\ref{eq:26}) can be derived by considering
Eq.~(\ref{eq:landaulifshitz-stereographic}) with a traveling wave
solution in the form
\begin{align}\label{eq:w-constantfrequency}
w(x_{1},x_{2},t)&=f(x_{1}-V\,t,x_{2})\,e^{i\,\omega\,t}\nonumber\\
&=f(\xi_1,\xi_{2})\,e^{i\,\omega\,t}\,.
\end{align}
Inserting the ansatz (\ref{eq:w-constantfrequency}) into Eq.
(\ref{eq:landaulifshitz-stereographic}) leads to the following
nonlinear eigenvalue problem
\begin{multline}
  \label{eq:f-equation}
    \omega\,f + i\,V f_{\xi_1} + f_{\xi_1\xi_1} + f_{\xi_{2}\xi_{2}}\\
    - \frac{2 f^*(f_{\xi_1}^2 + f_{\xi_{2}}^2)}{1 + |f|^2} - f \frac{1
      - |f|^2}{1 + |f|^2} = 0\,,
\end{multline}
with $f \to 0$ as $|\xiv| \to \infty$. Letting
\begin{equation*}
  f(\xi_1,\xi_{2})=u(\xi_1,\xi_{2})\,e^{-i\,\frac{V}{2}\,\xi_1}\,,
\end{equation*}
then (\ref{eq:f-equation}) becomes
\begin{equation}
  \label{eq:33}
  \begin{split}
    \omega\,u &+ \frac{V^2}{4} u + u_{\xi_1\xi_1} + u_{\xi_{2}\xi_{2}} \\
    &-
    \frac{2 u^*[(u_{\xi_1} - i \frac{V}{2} u)^2 + u_{\xi_{2}}^2]}{1 +
      |u|^2} - u \frac{1 - |u|^2}{1 +
      |u|^2} = 0\,.
  \end{split}
\end{equation}
We observe that for $V \ne 0$, $u$ is generally complex-valued.
Assuming localization and linearizing Eq.~(\ref{eq:33}) for $|\xiv|
\to \infty$, we obtain
\begin{equation*}
  u_{\xi_1\xi_1} + u_{\xi_{2}\xi_{2}} + \left ( \omega + \frac{V^2}{4}
    - 1 \right ) u = 0\,.
\end{equation*}
Therefore, a necessary condition for localization is
\begin{equation}
 \label{eq:36}
 \nu^2(V,\omega) > 0, \quad \nu(V,\omega) = \left (1 - \omega -
   \frac{V^2}{4} \right )^{1/2} \, ,
\end{equation}
and the exponential decay rate of $u$ is $\nu(V,\omega)$.  We discuss
the consequences of this observation in the next section.

\subsection{The nonlinear dispersion
  relation\label{sec:movingdroplet:nonlineardispersionrelation}}%

From the boundary condition in (\ref{eq:26}), we observe that the
propagating droplet is modulated by a spin wave background with wave
number
\begin{equation}
\label{eq:droplet-wavenumber} k=\frac{V}{2}\,,
\end{equation}
and frequency in the laboratory frame
\begin{equation}
  \label{eq:droplet-frequency} \wnl(k,\omega) =\omega+\frac{V^{2}}{2} = \omega +
  2k^2 \,.
\end{equation}
In contrast to the linear dispersion relation for exchange spin waves
in (\ref{eq:dispersionrelation-linear}), the nonlinear droplet
dispersion relation in (\ref{eq:droplet-frequency}) depends on both
the rest frequency $\omega$ and the speed $V$.  The condition
(\ref{eq:36}) corresponds to a negative nonlinear frequency shift
\begin{equation*}
  \wl(k) - \wnl(k,\omega) = 1 - \omega - k^2 = \nu^2(V,\omega) > 0 .
\end{equation*}
This yields an upper bound on the droplet's speed for a given rest
frequency
\begin{equation*}
  0 < V < \Vm(\omega) = 2 \sqrt{1 - \omega} ,
\end{equation*}
or, equivalently, an upper bound on the rest frequency for fixed speed
\begin{equation}
  \label{eq:34}
  \omega < \omega_\textrm{max}(V) = 1 - \frac{V^2}{4}\,, \quad V > 0 .
\end{equation}

In the limit $\omega \rightarrow 1$, condition (\ref{eq:34}), via
(\ref{eq:droplet-wavenumber}), requires that $k\rightarrow 0$.  The
nonlinear dispersion relation (\ref{eq:droplet-frequency}) then
coincides with the linear dispersion relation
(\ref{eq:dispersionrelation-linear}) for a spatially uniform, small
amplitude excitation at the bottom of the spin wave band.
Alternatively, in the limit $k\rightarrow
k_{\mathrm{max}}\equiv\sqrt{1-\omega}$, we observe that
$\omega_{NL}(k,\omega)\to 2-\omega=\omega_{L}(k_{\mathrm{max}})$
corresponding to a propagating exchange spin wave.  The speed $V$ of
the moving droplet is the group velocity $V_{g}$ of the background
spin wave, $V_{g} = \mathrm{d}\,\wl/\mathrm{d}\,k = 2\,k = V$.  The
phase speed of the background spin wave $\wl/k = \omega/k + 2k$ can be
greater than (when $\omega > 0$) or less than (when $\omega < 0$) the
droplet traveling wave speed $V = 2k$.  In this work, we focus on the
case $\omega > 0$.

\begin{figure}[!htb]
  \centering
  \includegraphics[width=0.99\columnwidth]{
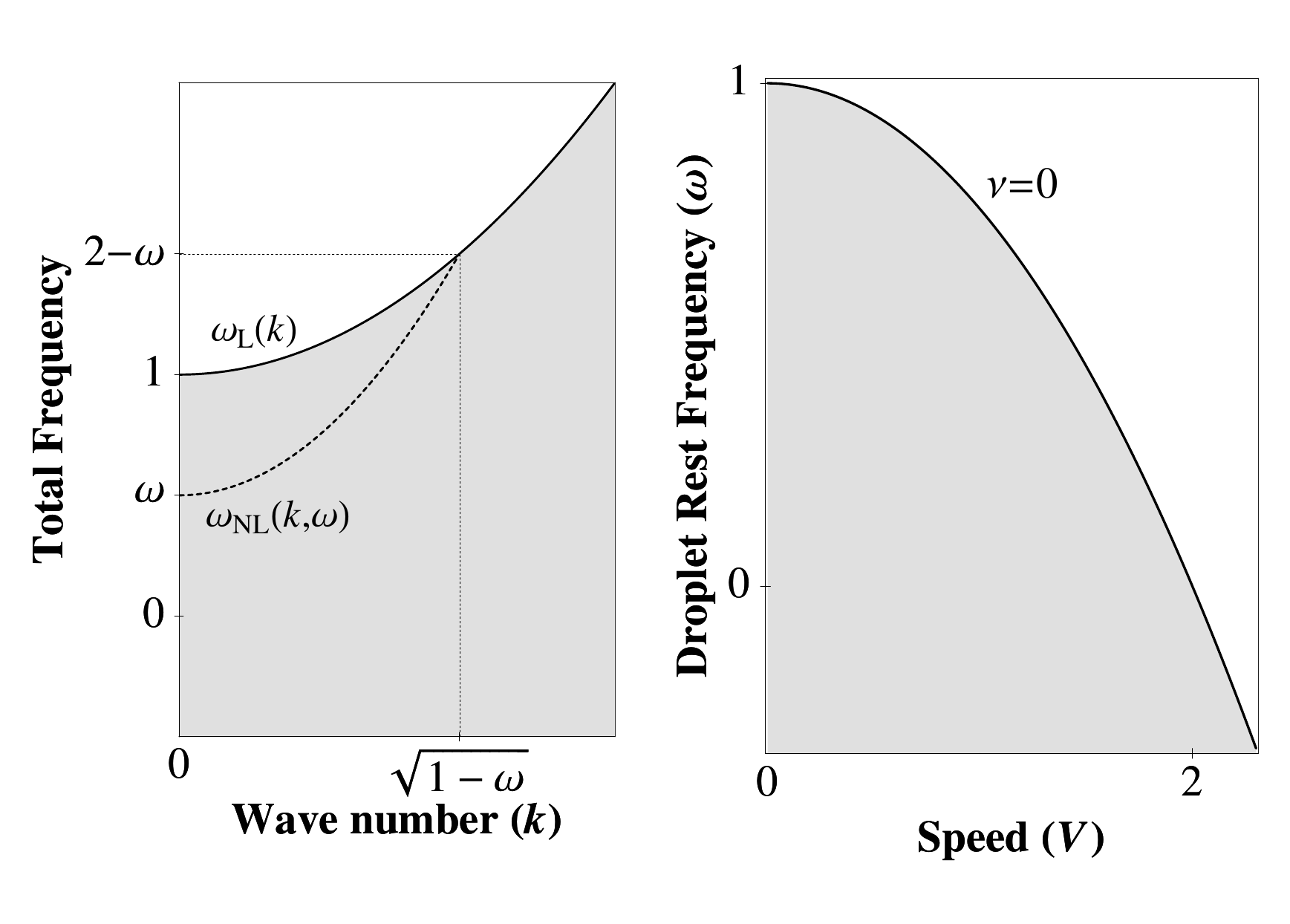}
  \caption{Dispersion relations in the laboratory frame
    (left) and the comoving droplet frame (right) with wavenumber $k =
    V/2$.  The shaded regions represent accessible droplet
    solutions. \label{fig:dispersionrelations}}
\end{figure}

The dispersion relations are illustrated in the laboratory frame and
the comoving droplet frame in Fig.~\ref{fig:dispersionrelations}.  The
shaded regions correspond to the localization condition (\ref{eq:36})
hence localized, propagating droplets can be interpreted as nonlinear
bound states lying below the linear spin wave band
\cite{kosevich_magnetic_1990}.

\subsection{1D Droplet\label{sec:1D droplet}}

The nonlinear dispersion relation in Eq.~(\ref{eq:droplet-frequency})
for a 2D propagating droplet is the same for a propagating 1D droplet
\cite{kosevich_magnetic_1990}.  As will be shown, a number of 2D
droplet properties are qualitatively similar to the 1D droplet
properties.  Therefore, in this section we briefly review some basic
results about moving droplet solutions of the LL equation in (1+1)
dimensions.


The Landau-Lifshitz Eq.~(\ref{eq:landaulifshitz-vector}) in one
dimension, $\Delta \to \partial_{xx}$, is solvable by means of the
inverse scattering transform (see, \textit{e.g.},
\cite{chen_huang_liu_1995} and references therein) and admits soliton
solutions.  The single droplet soliton solution can be written as
follows (see \cite{long_nonlinear_1979})
\begin{subequations}
  \label{eq:1D solution}
  \begin{align}
    \cos \Theta = &~\frac{V^2 - 2 + 3 \omega + \sqrt{V^2 + \omega^2}
      \cosh[2 \nu \xi]}{2 - \omega + \sqrt{V^2 + \omega^2} \cosh [ 2
      \nu \xi]} \\
    \nonumber
    \Phi = &~\omega t -\frac{V\xi}{2} \\
    \label{eq:1}
    &+ \tan^{-1} \left [\frac{4 \nu
        \tanh(\nu\xi/2)}{V^2 + 2(\omega + \sqrt{V^2 + \omega^2})} \right
    ]\, ,
  \end{align}
\end{subequations}
where $\xi=x-V\,t$ and $\nu$ is given in (\ref{eq:36}). 
Thus, the 1D soliton solution exists for all $V > 0$, $\omega$
satisfying (\ref{eq:36}).  When $V = 0$, the rest frequency is limited
to the unit interval (\ref{eq:omega-interval}).  For $\omega < 0$, $V
= 0$, the soliton (\ref{eq:1D solution}) exhibits a phase singularity
at the origin corresponding to topological behavior in this one
dimensional setting.  However, for $V > 0$, the necessary condition
$\omega < \omega_{\mathrm{max}}(V)$ in (\ref{eq:34}) has no lower
bound.  Therefore, nontopological solitons with arbitrarily negative
rest frequencies exist even with vanishingly small propagation speeds.
The phase gradient (\ref{eq:1}) has the asymptotics
\begin{equation*}
  \Phi_\xi(0) \sim \frac{-V}{\omega + | \omega| + V^2/(2|\omega|)},
  \quad 0 < \frac{V}{|\omega|} \ll 1,
\end{equation*}
hence can be quite large when $\omega < 0$.  Our numerical method for
computing 2D propagating droplets fails for modes with large momentum
due to large phase gradients (see Sec.~\ref{sec:droplet properties})
so we are limited in exploring this regime of parameter space in the
two dimensional case.

Explicit expressions for the energy ${\cal{E}}^{(\mathrm{1D})}$, the
spin density ${\cal{N}}^{(\mathrm{1D})}$ and the momentum
${\cal{P}}^{(\mathrm{1D})}$, as functions of $V$ and $\omega$, for a
single moving droplet solution, can be obtained from solution
(\ref{eq:1D solution}) via Eqs.~(\ref{eq:3}), (\ref{eq:4}) and
(\ref{eq:5}): 
\begin{align*}
  {\cal{E}}^{(\mathrm{1D})} &=2\,\sqrt{4-4\,\omega-V^{2}}\,,\\%
  {\cal{N}}^{(\mathrm{1D})} &=2\,\tanh^{-1}{\left(\frac{\sqrt{4-4\,\omega-V^{2}}}{2-\omega}\right)}\,,\\%
  {\cal{P}}^{(\mathrm{1D})} &=\pi-2\,\tan^{-1}{\left(\frac{V^{2}+2\,\omega}{V\,\sqrt{4-4\,\omega-V^{2}}}\right)}\,.%
\end{align*}
We observe the notable fact that the momentum
${\cal{P}}^{(\mathrm{1D})}$ is bounded, $0 \le
{\cal{P}}^{(\mathrm{1D})}<2\,\pi$, but the energy and spin density are
not.

\section{Propagating Droplet:  Asymptotic Results}
\label{sec:prop-dropl-asympt}

In this section, asymptotic methods are employed to study the behavior
of the propagating droplet in the small amplitude and small speed
regimes.

\subsection{Weakly nonlinear regime} \label{sec:small-amplitude}

Small amplitude, propagating droplets were studied asymptotically in
\cite{ivanov_zaspel_yastremsky_2001}.  In this section, we briefly
review the results with a view toward their numerical verification in
Sec.~\ref{sec:moving-dropl-numer}.

Using a slowly varying envelope, small amplitude approximation the
authors in \cite{ivanov_zaspel_yastremsky_2001} construct a
propagating droplet bifurcating from the set of linear waves where the
small parameter is the deviation from the linear band edge
(\ref{eq:36})
\begin{equation*}
  0 < \nu(V,\omega)^2 \ll 1 \, .
\end{equation*}
The droplet solution for Eq.~(\ref{eq:landaulifshitz-stereographic})
takes the approximate form
\begin{equation}
  \label{eq:41}
  \begin{split}
    w(x_{1},x_{2},t) &\sim \frac{\nu(V,\omega)}{\sqrt{2(2-\omega)}} e^{ -i
      \left [ V(x_{1} - V\,t)/2 - \omega t \right ]}
    \\
    \times u_T &\left (
      \nu(V,\omega) \left [(x_{1} - V\,t)^2 +
        x_{2}^2\right
      ]^{1/2} \right )\,, 
  \end{split}
\end{equation}
with $\nu > 0$ having the meaning of the inverse half width.  As the
band edge is approached, $\nu \to 0$, the propagating droplet is
approximately a wide, small amplitude, propagating Townes mode
\cite{chiao_trapping_1964}, the unique, positive, monotonically
decaying solution of
\begin{equation*}
  \begin{split}
    - u_T'' - &\frac{1}{R} u_T' - u_T^3 + u_T = 0 \, , \\
    u_T'(0) &= 0, \quad \lim_{R \to \infty} u_T(R) = 0\, .
  \end{split}
\end{equation*}

Using (\ref{eq:41}) and its next order correction, the
functional (\ref{eq:27}) can be expanded as
\cite{ivanov_zaspel_yastremsky_2001}
\begin{equation}
  \label{eq:38}
  \mathcal{D} \sim \frac{\mathcal{N}_T}{1 + V^2/4} \left [ \nu^2 +
    \nu^4 \frac{b(1 - V^2/4)}{2 (1 + V^2/4)^2} \right ] \, ,
\end{equation}
where
\begin{equation*}
  \begin{split}
    \mathcal{N}_T &\equiv \| u_T \|_{L^2(\R^2)}^2 \approx 11.7009
    \, , \\
    b &\equiv \frac{2\pi \int_0^\infty u_T^2 {u_T'}^2 R\,
      \mathrm{d} R}{\mathcal{N}_T} \approx 1.35825 \, .
  \end{split}
\end{equation*}
Combining (\ref{eq:38}) with the relations (\ref{eq:37}), the
propagating droplet's spin density, momentum, and energy in the
weakly nonlinear regime are approximately
\begin{subequations}
  \label{eq:48}
  \begin{align}
    \label{eq:47}
    \mathcal{N} &\sim \frac{\mathcal{N}_T}{1 + V^2/4}\left [ 1
      + \nu^2 \frac{b (1 - V^2/4)}{(1 + V^2/4)^2} \right ] \\
    \label{eq:51}
    \mathcal{P}_1 &\sim \frac{\mathcal{N}_T V}{2(1 + V^2/4)}
    \bigg [  1 + \nu^2 \bigg ( b \frac{1 - V^2/4}{(1 + V^2/4)^2}
    \\
    \nonumber 
    &\qquad \qquad \qquad \qquad \quad +
    \frac{1}{1 + V^2/4} \bigg )\bigg] \, ,\\
    \label{eq:52}
    \mathcal{E} 
    &\sim \mathcal{N}_T \bigg [ 1 + \nu^2 \bigg (
    \frac{2 b (1 - V^2/4)}{(1 - V^2/4)^3} \\
    \nonumber & \qquad \qquad \qquad  + \frac{1 + 3V^2/4}{(1 + V^2/4)^2} -
    \frac{1}{1 + V^2/4}
    \bigg )  \bigg ] \, .
  \end{align}
\end{subequations}
These results are numerically verified in Sec.~\ref{sec:droplet
  properties}.

For the initial value problem in the weakly nonlinear regime, one can
derive at leading order the time-dependent cubic Nonlinear
Schr\"{o}dinger equation (NLS) from
Eq.~(\ref{eq:landaulifshitz-stereographic}) for the slowly varying
envelope of a carrier wave.  In two spatial dimensions, the NLS
equation exhibits critical blow up with the Townes mode playing an
important role \cite{sulem_nonlinear_1999}.  For initial data with
$L^2(\R^2)$ norm less than $\| u_T \|_2$, the solution disperses to
zero as $t \to \infty$.  In other cases, e.g.~for data close to $u_T$
in $L^2(\R^2)$ but with larger norm, the solution blows up in finite
time.  Thus, the Townes mode is unstable to small perturbations.  It
is natural then to inquire into the stability of the propagating
droplet for the LL Eq.~(\ref{eq:landaulifshitz-stereographic}).  The
Jacobian condition (\ref{eq:88}) was verified in
\cite{ivanov_zaspel_yastremsky_2001}.  Insertion of (\ref{eq:47}) and
(\ref{eq:51}) into Eq.~(\ref{eq:83}), leads to
\begin{equation}
  \label{eq:84}
  J = -\frac{b \omega \mathcal{N}_T}{2(2 - \omega)^4} +
  \mathcal{O}(\nu^2)\, ,
\end{equation}
suggesting that the higher order corrections to the NLS approximation
stabilize the propagating droplet.  Further discussion of stability in
the large amplitude case is presented in
Sec.~\ref{sec:droplet-stability}.


\subsection{Small propagation
  speed \label{sec:movingdroplet:smallvelocity}}%

We now consider Eqs.~\eqref{eq:20} and \eqref{eq:24} with boundary
conditions (\ref{eq:22}) and (\ref{eq:26}) for small values of the
droplet speed $0 < V\ll 1$.  Polar coordinates are used (see
Fig.~\ref{fig:refsys})
\begin{equation}
  \label{eq:polarcoordinates-movingframe}
  \xi_1 = x_1-V t=\rho\,\cos{\varphi} \,,\quad \xi_2 = x_2 =
  \rho\,\sin{\varphi} \,.
\end{equation}
A solution is sought in the form
\begin{equation}
  \label{eq:asymptotics-smallV}
  \begin{split}
    \theta& = \Theta_{0} (\rho;\omega) +\mathcal{O}(V^{2})\,,\\
    \phi& = V\, \left ( \Phi_{1}(\rho,\varphi) - \frac{1}{2}
      \,\rho\,\cos{\varphi}\right ) +\mathcal{O}(V^{2})\,,
  \end{split}
\end{equation}
where $\Theta_{0}$ is the solution of Eq.~(\ref{eq:Theta0-equation})
for the stationary droplet with $0 < \omega < 1$ and $\theta$ and
$\phi$ are given in the ansatz (\ref{eq:8}).  The azimuthal angle
$\phi$ in Eq.~\eqref{eq:asymptotics-smallV} should not be confused
with the angle $\varphi$ of
Eq.~\eqref{eq:polarcoordinates-movingframe}; see
Fig.~\ref{fig:refsys}.

At first order in $V$, the function $\Phi_{1}$ satisfies
\begin{equation}
  \label{eq:Phi1}
  \sin{\Theta_{0}}\,\Delta\Phi_{1}+
  2\,\Theta_{0}'\,\frac{\partial\Phi_{1}}{\partial\rho}\,\cos{\Theta_{0}}
  = (\cos \Theta_0 - 1) \Theta_{0}'\,\cos{\varphi}\,,
\end{equation}
subject to the boundary condition
\begin{equation*}
  \lim_{\rho \to \infty} \nabla \Phi_1 = 0 .
\end{equation*}
Here and hereafter the prime indicates derivation with respect to
$\rho$.  Equation (\ref{eq:Phi1}) admits separation of variables in
the form
\begin{equation*}
  \Phi_{1}(\rho,\varphi) =
  \frac{\tilde{\Phi}_{1}(\rho)\,\cos{\varphi}}{\sin{\Theta_{0}}(\rho)}\,,
\end{equation*}
where $\tilde{\Phi}_{1}$ is the solution of the ordinary differential
equation (ODE)
\begin{subequations}
\begin{align}
  \label{eq:Phi1tilde-equation}
  L_{\Phi} \, \tilde{\Phi}_{1}  &= (1 - \cos \Theta_0) \Theta_{0}'\,,
  \\
  L_{\Phi} &= -\frac{\mathrm{d}^2}{\mathrm{d} \rho^2} -
  \frac{1}{\rho} \frac{\mathrm{d}}{\mathrm{d}\rho} +\frac{1}{\rho^{2}}
  + V_\Phi(\rho) \,
  ,\\
  V_\Phi(\rho) &=  - {\Theta_{0}'}^{2}
  + \left(\cos{\Theta_{0}}-\omega\right)\, \cos{\Theta_{0}} \, ,
\end{align}
subject to
\begin{equation}
  \label{eq:67}
  \lim_{\rho \to \infty}
    \frac{\mathrm{d}}{\mathrm{d}\,\rho}\,\left[\sqrt{r}\,e^{\rho \sqrt{1 - \omega}}
  \tilde{\Phi}_1(\rho)\right]  = 0\,,\quad \tilde{\Phi}_1(0) = 0\, ,
\end{equation}
\end{subequations}
from the asymptotics of the stationary droplet (\ref{eq:68}).


Equation (\ref{eq:Phi1tilde-equation}) is solvable for any choice of
the parameter $0<\omega<1$. This can be seen as follows. Standard
arguments (\textit{e.g.} using variational methods) show that
$L_{\Phi}$ is a positive definite operator. The solution to
(\ref{eq:Phi1tilde-equation}),
\begin{equation*}
  \tilde{\Phi}_1 = L_\Phi^{-1} \left \{ (1 - \cos \Theta_0) \Theta_0'
  \right \} \, ,
\end{equation*}
satisfies the boundary condition (\ref{eq:67}) as shown by the
asymptotics,
\begin{equation*}
  \begin{split}
    \tilde{\Phi}_{1}&=\mathcal{O}\left(
      \frac{e^{-\rho\,\sqrt{1-\omega}}}{\sqrt{\rho}}\right)\quad  
    \rho\gg1\,,\\ 
    \tilde{\Phi}_{1}&=\mathcal{O}\left(\rho\right)\quad 0<\rho\ll1\,.
  \end{split}
\end{equation*}
Therefore, from the ansatz (\ref{eq:8}) and
(\ref{eq:asymptotics-smallV}), we find that an approximate droplet
solution for small $V$ consists of the stationary solution modulated
by the spatially varying phase
\begin{equation*}
  \Phi = \omega t - \frac{V}{2}
  x_1+V\,\Phi_{1}(\rho,\varphi)+\mathcal{O}\left(V^{2}\right) \, .
\end{equation*}
\section{Propagating Droplet: Numerical Results \label{sec:moving-dropl-numer}}%

In this section, a numerical method is introduced to solve
Eq.~(\ref{eq:f-equation}) for propagating droplets.  The solutions'
properties are investigated and its stability verified by the
analytical criterion $J > 0$ and time dependent numerical simulations.

\subsection{Iterative Method to Compute Propagating
  Droplets \label{sec:numerical solution BVP}}%

A solution to the nonlinear eigenvalue problem (\ref{eq:f-equation})
is computed by the spectral renormalization method
\cite{ablowitz_spectral_2005}, a generalization of the Petviashvili
method \cite{petviashvili_equation_1976}.

We begin the procedure by rephrasing Eq.~(\ref{eq:f-equation}) as
follows
\begin{subequations}
 \label{eq:bvp-equation-LNoperators}
\begin{equation}
\omega f + i V f_{x_{1}} + \Delta f =N[f]\,,
\end{equation}
with
\begin{align}
&N[f] =\frac{2 f^* (f_{x_{1}}^2 + f_{x_{2}}^2)+f\,(1 - |f|^2)}{1 +
|f|^2}\,,
\end{align}
\end{subequations}
and, upon taking its Fourier transform, results in
\begin{equation*}
\left(\omega-V\,\kappa_{1}-\kappa^{2}\right)\,\widehat{f}=\widehat{N[f]}
\, ,
\end{equation*}
where $\widehat{f}=\mathcal{F}[f](\vec{\kappa})$ is the Fourier
transform of $f$ and $\vec{\kappa}=(\kappa_{1},\kappa_{2})$ is the
Fourier wavenumber, with $|\vec{\kappa}|=\kappa$.  The
iteration
\begin{equation}
(\omega-V\,\kappa_{1}-\kappa^{2}-c)\,\widehat{F}_{n+1}=
\widehat{N[f_{n}]}-c\,\widehat{f}_{n} \,,
\end{equation}
is introduced so that
\begin{equation*}
\widehat{F}_{n+1}=\frac{\widehat{N[f_{n}]}-c\,\widehat{f}_{n}}
{\omega-c- V\,\kappa_{1}-\kappa^{2}}\,,
\end{equation*}
where $c$ is a constant real parameter, chosen in order to avoid
divergence of the iteration -- its value is experimentally obtained,
see \cite{ablowitz_spectral_2005}.  The value of $c$ affects the
convergence speed of the iteration and must be taken larger as the
droplet amplitude is increased.  In most of our computations, we find
$c = 10$ to suffice.

In order to iterate the sequence $\{f_{n}\}$, we scale $F_{n+1}$
according to
\begin{equation}
 \label{eq:bvp-equation-iteration-b}
f_{n+1}=\eta_{n+1}\,F_{n+1}\,,
\end{equation}
and choose the complex parameter $\eta_{n+1}$ so that the projection
of Eq.~(\ref{eq:bvp-equation-LNoperators}) onto $f_{n+1}$ is
satisfied. Inserting (\ref{eq:bvp-equation-iteration-b}) into Eq.
(\ref{eq:bvp-equation-LNoperators}) and taking the $L^2(\R^2)$ inner
product with $f_{n+1}$ results in
\begin{gather}
  \label{eq:bvp-equation-integrals}
  \int\bigg[\omega\,|F_{n+1}|^{2}+i\,V\,F_{n+1}^{*}\,\frac{\partial
    F_{n+1}}{\partial x}\\
  \nonumber
  -|\nabla\,F_{n+1}|^{2}\bigg]\,\mathrm{d}x_{1}\,\mathrm{d}x_{2}=
  |\eta_{n+1}|^{2}\,\times\\
  \nonumber
  \Bigg\{2\,\int\frac{
    {\left(F_{n+1}^{{*}}\right)}^{2}
    \,\left[\left(\frac{\partial F_{n+1}}{\partial x}\right)^{2} +
      \left(\frac{\partial F_{n+1}}{\partial y}\right)^{2}\right]}
  {1+|\eta_{n+1}|^{2}\,|F_{n+1}|^{2}}\,\mathrm{d}x_{1}\,\mathrm{d}x_{2}\,\\
  \nonumber
  +\int\frac{|F_{n+1}|^{2}\,
    \left(1-|\eta_{n+1}|^{2}\,|F_{n+1}|^{2}\right)}
  {1+|\eta_{n+1}|^{2}\,|F_{n+1}|^{2}}\,\mathrm{d}x_{1}\,\mathrm{d}x_{2}\,\Bigg\}\,.
\end{gather}
Equation (\ref{eq:bvp-equation-integrals}) determines $|\eta_{n+1}|$.

The numerical computations are implemented on the finite, discrete
grid
\begin{equation*}
    {x_{1,2}}_{j} = -L + (j-1)\,\Delta\,, \,\, j = 1, 2, \ldots, N\,,
\end{equation*}
with
\begin{equation*}
  \Delta=\frac{2L}{N}\,. 
\end{equation*}
Due to the localization of $F_{n+1}$, derivatives are estimated using
the fast Fourier transform.  We take $L = 50$ and $N = 1536$, implying
$\Delta\approx0.065$, unless otherwise stated. To obtain droplets with
large speed, we use continuation in $V$. For small $V$, the initial
guess $f_0$ is the stationary droplet discussed in
Sec.~\ref{sec:stationarydroplet} with the same $\omega$.  For large
$V$, $f_0$ is a previously computed propagating droplet with the same
$\omega$ but smaller $V$. The integrals in
(\ref{eq:bvp-equation-integrals}) are computed using the spectrally
accurate trapezoidal rule.

Equation (\ref{eq:bvp-equation-integrals}) is solved for
$|\eta_{n+1}|^{2}$ using Newton's method and the phase of $\eta_{n+1}$
is determined by the condition that $f_{n+1}$ have a vanishing phase
at the origin so that
\begin{equation}
f_{n+1}=|\eta_{n+1}| \,F_{n+1}\,e^{-i\,[\arg{F_{n+1}(0,0)}]}\,,
\end{equation}
This implies that the projection of the magnetization at the origin
onto the plane points along $e_1$.

\begin{figure*}[htb!]
  \centering
  \includegraphics[scale=1]{
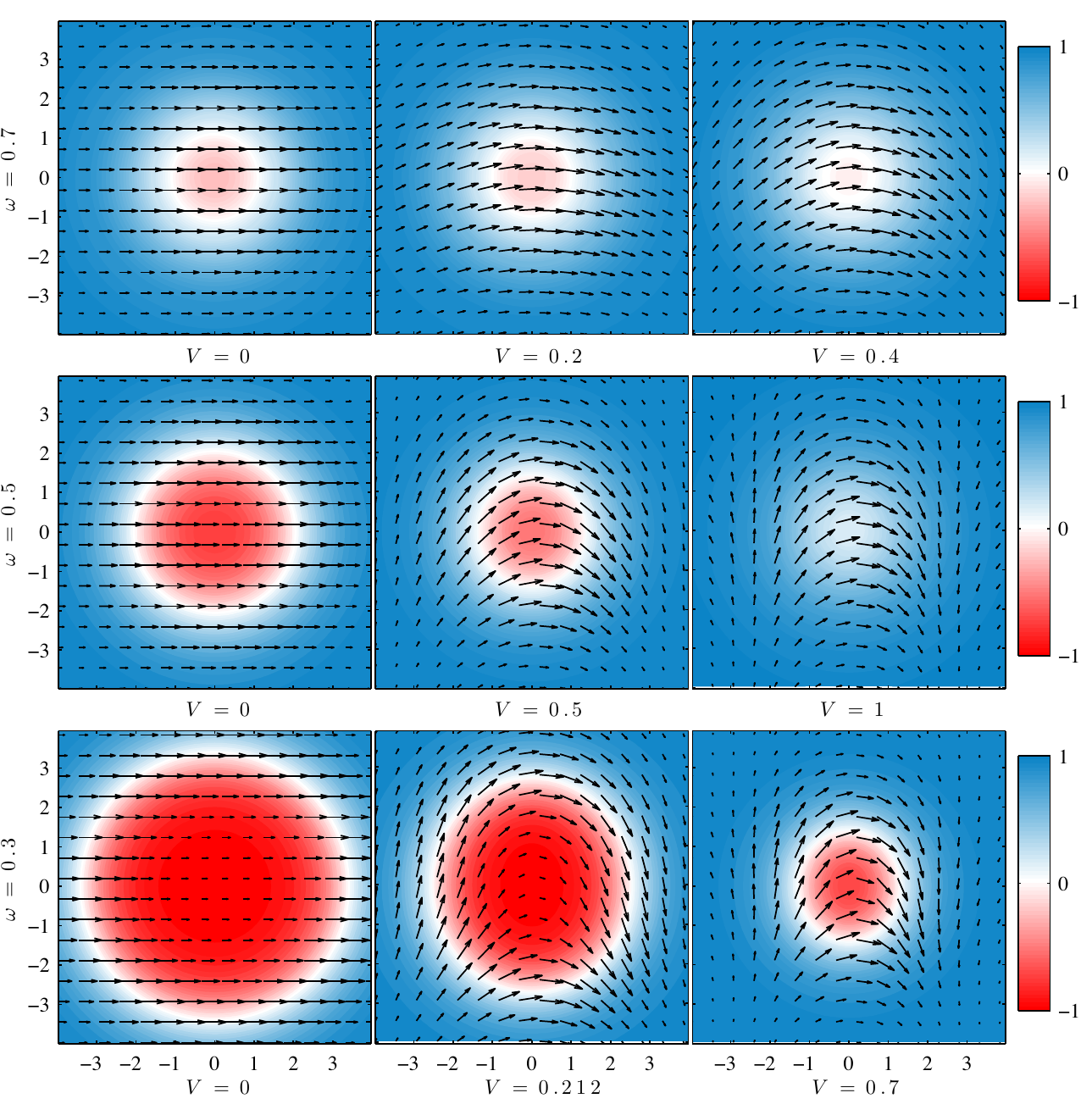}
  \caption{Computed droplet solutions for several choices of $\omega$
    and $V$. The color contour plot represents the magnitude of
    $m_{3}$ and the arrows represent the projection of the
    magnetization on the plane.  To facilitate visualization, the
    length of the projected field is normalized to the largest value
    in each frame.  \label{fig:numericalsolution:bvp}}
\end{figure*}
\begin{figure*}[htb!]
  \centering
  \includegraphics[scale=1]{
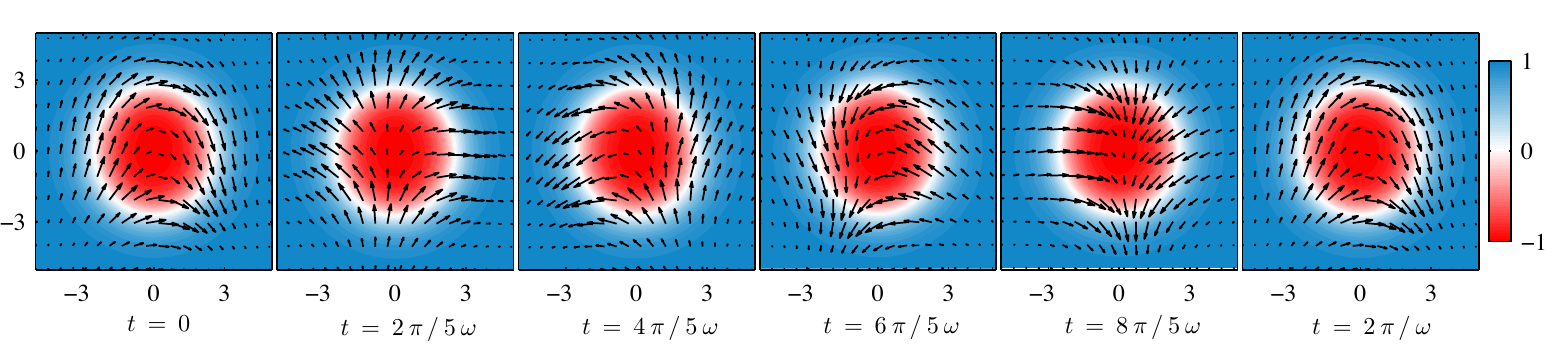}
  \caption{Precessional motion of a propagating droplet with
    $\omega=0.3$ and $V=0.212$ in the comoving frame.  Each frame depicts
    a different instant in time over one period
    $2\pi/\omega$. \label{fig:droplet_evolution}}
\end{figure*}

It is helpful to introduce the residual
\begin{equation*}
  r[f_n] = \mathcal{F}^{-1} \left [ \widehat{N f_n} - \left (
      \omega - V \kappa_1 - \kappa^2 \right ) \widehat{f} \right ],
\end{equation*}
which is zero at a droplet solution.  We iterate from $f_{0}$ until
\begin{align*}
  &\|f_{n}-f_{n-1}\|_\infty <\mathrm{tol}_{f}\,,\\
  &\|r[f_{n}] \|_\infty < \mathrm{tol}_r\,,\\
  &|\tilde{\eta}_{n}-\tilde{\eta}_{n-1}| <\mathrm{tol}_{\eta}\,,
\end{align*}
where we set $\mathrm{tol}_{f}=\mathrm{tol}_{\eta}=\mathrm{tol}_r
=10^{-9}$, unless otherwise stated.

Further details and validation of this numerical scheme are presented
in the Appendix.

\begin{figure*}[!htb]
  \centering
  \subfigure[{Energy, $\mathcal{E}$}\label{fig:contourplots-energy}]%
  {{\includegraphics[scale=1]{
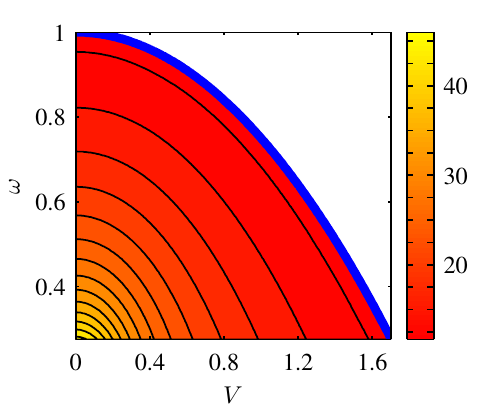}}}\quad%
  \subfigure[{Momentum, $\mathcal{P}_{1}$}\label{fig:contourplots-momentum}]%
  {{\includegraphics[scale=1]{
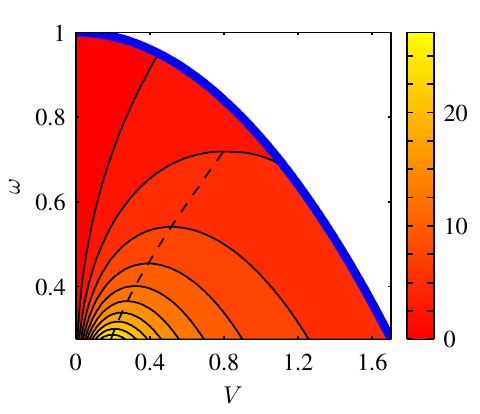}}}\quad%
  \subfigure[{Spin density, $\mathcal{N}$}\label{fig:contourplots-spindensity}]%
  {{\includegraphics[scale=1]{
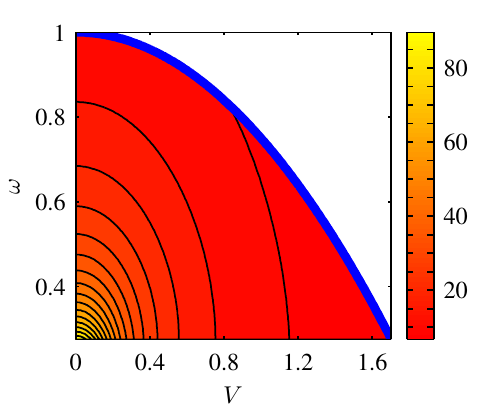}}}\\%
  \subfigure[{Amplitude: $1-\min{(m_{3})}$}\label{fig:contourplots-amplitude}]%
  {{\includegraphics[scale=1]{
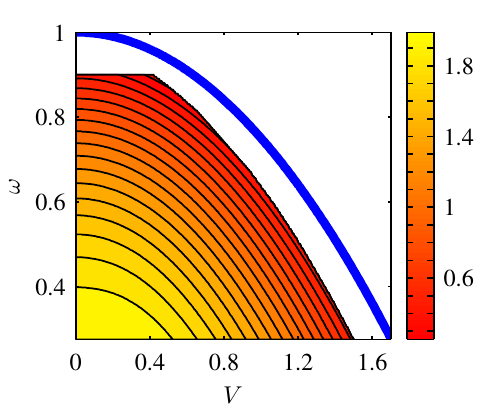}}}\quad%
  \subfigure[{Width ($x_1$-direction): $\mbox{width}_{1}$}\label{fig:contourplots-widthx}]%
  {{\includegraphics[scale=1]{
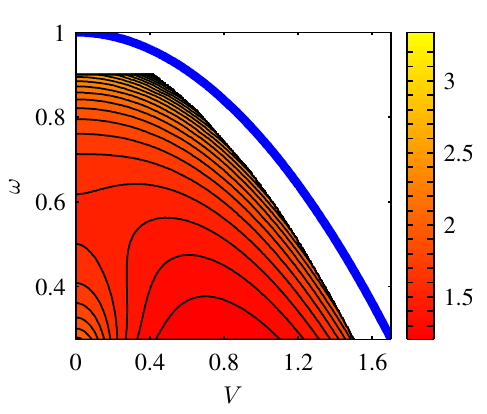}}}\quad%
  \subfigure[{Asymmetry ratio: $\frac{\mbox{width}_{2}-\mbox{width}_{1}}{\mbox{width}_{1}}$}\label{fig:contourplots-widthratio}]%
  {{\includegraphics[scale=1]{
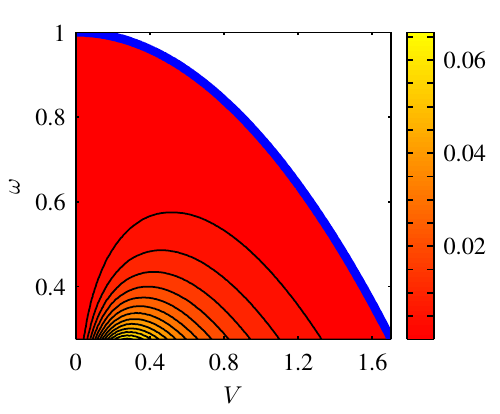}}}%
  \caption{Contour plots of the droplet's energy, momentum, spin density, amplitude,
  width in the $x$-direction and asymmetry ratio, computed from numerical solutions
  of the boundary value problem. \label{fig:contourplots}}
\end{figure*}

\subsection{Droplet Properties\label{sec:droplet properties}}%

Using the numerical scheme presented in the previous section, we
computed a large number of solutions to the boundary value problem
(\ref{eq:f-equation}) for a propagating droplet with parameters lying
in the set
\begin{equation*}
  \begin{split}
    S = \big \{ (V,\omega) ~ | ~
      &0 \le V \le 2 \sqrt{1 - \omega} - 0.25 ,\\
      &0.275 \le \omega \le 0.95 \big \} .
  \end{split}
\end{equation*}
This representative set of droplet parameters was chosen because
accurate solutions were obtained without having to change the
numerical parameters of our scheme.  Outside of this set, either
higher resolution or larger domains were often required to accurately
resolve the droplets (see Appendix).  Droplets with $\omega < 0$ were
computed and exhibit similar properties to those in $S$.  We were
unable to compute large momentum droplets for particular $V$ with
$\omega$ sufficiently small or negative due to inherently large phase
gradients.  Illustrative example droplet solutions are shown in
Fig.~\ref{fig:numericalsolution:bvp}.  As $\omega$ and $V$ are
increased, the droplet amplitude, defined as $1 - \min{(m_{3})} = 1 -
m_3(0,0)$, decreases.  The spin wave background is evident, exhibiting
a decreasing wavelength as $V$ is increased or as $\omega$ is
decreased.  This is in accord with the nonlinear dispersion relation
discussed in Sec.~\ref{sec:movingdroplet:nonlineardispersionrelation}.
The precessional motion of a propagating droplet is depicted in Fig.
\ref{fig:droplet_evolution} in the comoving frame.

Using Eq.~(\ref{eq:49}), the energy $\mathcal{E}$, momentum
$\mathcal{P}_{1}$, and spin density $\mathcal{N}$ were computed over
$S$.  Contour plots of these quantities are shown in Figs.
\ref{fig:contourplots-energy}-\ref{fig:contourplots-spindensity}.
Additional values of the conserved quantities outside $S$ were
obtained by interpolation using the weakly nonlinear asymptotics
(\ref{eq:48}).  Both the energy and the spin density are monotone
decreasing functions of $V$ and $\omega$.  On the contrary,
Fig.~\ref{fig:contourplots-momentum} shows that the momentum
dependence is not monotonic.  For fixed values of $\omega$, the
momentum $\mathcal{P}_{1}$ exhibits a maximum for $V > 0$.  This
``ridgeline'' is shown as the dashed curve in
Fig.~\ref{fig:contourplots-momentum}.  The droplets with the largest
momentum exhibit the largest phase gradient near the origin.  See, for
example, the case $V = 0.212$, $\omega = 0.3$ in
Figs.~\ref{fig:numericalsolution:bvp} and \ref{fig:droplet_evolution}
which is on the ridgeline of large momentum droplets.

\begin{figure*}[!htb]
  \centering
  \subfigure[{Energy, $\mathcal{E}$}\label{fig:weaklynl-energy}]%
  {{\includegraphics[scale=1]{
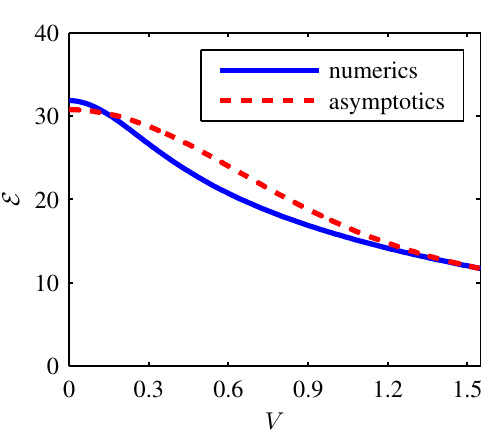}}}\quad%
  \subfigure[{Momentum, $\mathcal{P}_{1}$}\label{fig:weaklynl-momentum}]%
  {{\includegraphics[scale=1]{
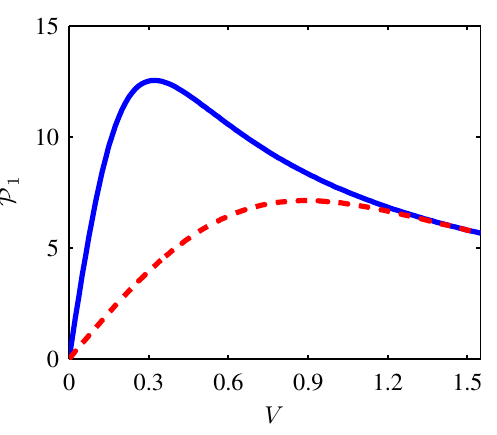}}}\quad%
  \subfigure[{Spin density, $\mathcal{N}$}\label{fig:weaklynl-spindensity}]%
  {{\includegraphics[scale=1]{
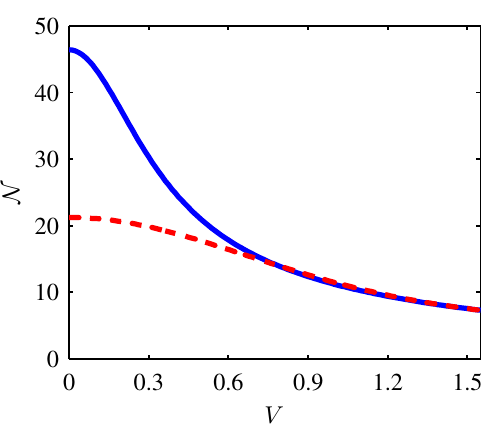}}}
\caption{Comparison of the weakly nonlinear asymptotics (\ref{eq:48})
  with numerical computation of $\mathcal{E}$, $\mathcal{P}$, and
  $\mathcal{N}$ for $\omega = 0.4$, $\Vm \approx
  1.55$. \label{fig:weakly_nl}}
\end{figure*}
Figure \ref{fig:weakly_nl} depicts the numerical verification of the
weakly nonlinear results (\ref{eq:48}) for fixed $\omega = 0.4$.  The
results agree near the band edge where $\Vm - V \ll 1$ and we have
verified that the error exhibits the expected $\nu^4$ scaling.

The droplet's amplitude dependence is shown in Fig.
\ref{fig:contourplots-amplitude}.  As observed earlier, larger
amplitude droplets correspond to smaller speeds and rest
frequencies.

The droplets shown in Figs.~\ref{fig:numericalsolution:bvp} and
\ref{fig:droplet_evolution} show negligible signs of asymmetry in
their localization extent.  To further investigate this, we define
the droplet widths in the $x_1$ and $x_2$ directions according to
the standard deviations
\begin{subequations}
  \begin{align}
    \label{eq:widths}
    \mbox{width}_{1}=&{\left(\frac{1}{\mathcal{N}}\, \int\,x_{1}^{2}\,
        \left(1-\cos{\Theta}\right)\,
        \mathrm{d}\vec{x}\right)}^{\frac{1}{2}}\,,\\
    \mbox{width}_{2}=&{\left(\frac{1}{\mathcal{N}}\,
        \int\,x_{2}^{2}\,\left(1-\cos{\Theta}\right)\,
        \mathrm{d}\vec{x}\right)}^{\frac{1}{2}}\,.
  \end{align}
\end{subequations}
A contour plot of the quantity $\mbox{width}_1$ for $(V,\omega) \in
S$ is shown in Fig.~\ref{fig:contourplots-widthx}, exhibiting non
monotonic behavior for increasing $V$ and $\omega$, especially for
large amplitude modes.  The droplet's width diverges according to
$\mathcal{O}(1/\nu(V,\omega))$ as the linear spin wave band is
approached $0 < \nu \ll 1$ (see Eq.~(\ref{eq:41})).

The asymmetry, defined as $(\mbox{width}_2 -
\mbox{width}_1)/\mbox{width}_1$, is a nonnegative quantity as shown
in Fig.~\ref{fig:contourplots-widthratio}.  Even though the
asymmetry does increase -- approximately with the momentum -- it is
very small for droplets with parameters in $S$, reaching a maximum
of approximately 0.066.  Wide, small amplitude droplets are
symmetric as shown in Eq.~(\ref{eq:41}).

We now comment briefly on the localized excitations of Equation
\eqref{eq:landaulifshitz-vector} observed in
\cite{piette_localized_1998}.  These states were formed by starting
with a stationary droplet initial condition $f_0(\rho;\omega) = \sin
\Theta_0(\rho;\omega)/[1 + \cos \Theta_0(\rho;\omega)]$ to
Eq.~\eqref{eq:landaulifshitz-stereographic} superimposed with a
spatially varying phase in the form
\begin{equation}
  \label{eq:9}
  w(\x,t=0) = e^{i \pi \kv \cdot \x} f_0(|\x|;\omega) .
\end{equation}
This initial condition to
Eq.~\eqref{eq:landaulifshitz-stereographic} was allowed to evolve in
time and a moving, localized structure was observed.  The frequency
of precession $\omega$, speed $V$, and energy $\mathcal{E}'$
associated with several initial conditions reported in
\cite{piette_localized_1998} are given in Table \ref{tab:piette}.
For comparison, the energies $\mathcal{E}$ of directly computed
propagating droplets with the same frequency and speed are also
listed in Table \ref{tab:piette}.  The droplet approximation
\eqref{eq:9} leads to overestimated energies which are less accurate
for larger speeds.  This behavior coincides with the small $V$
asymptotics of Sec.~\ref{sec:movingdroplet:smallvelocity} which show
that the ansatz in Eq.~(\ref{eq:9}) is approximately valid.
\begin{table}[h]
  \centering
  \setlength{\tabcolsep}{10pt}
  \begin{tabular}{|c|c|c|c|}
    \hline
    $V$ & $\omega$ & $\mathcal{E}'$ & $\mathcal{E}$ \\
    \hline
    0.067 & 0.50 & 25.6 & 25.5 \\
    0.13 & 0.49 & 25.9 & 25.8 \\
    0.24 & 0.47 & 27.0 & 26.9 \\
    0.3 & 0.46 & 28.9 & 27.8 \\
    0.35 & 0.44 & 31.4 & 28.8 \\
    \hline
  \end{tabular}
  \caption{Comparison between the energies
    $\mathcal{E}'$ computed  in \cite{piette_localized_1998}  by
    solving the initial value problem
    \eqref{eq:9} with the
    energies $\mathcal{E}$ of droplets computed in this work. }
  \label{tab:piette}
\end{table}



\subsection{Propagating Droplet Stability}
\label{sec:droplet-stability}

The nonlinear, orbital stability of droplets has been supported by
verification of the Jacobian condition (\ref{eq:87}) in the stationary
\cite{kosevich_magnetic_1990} and weakly nonlinear
\cite{ivanov_zaspel_yastremsky_2001} (recall Eq.~(\ref{eq:84})) cases.
In this section, we confirm the stability of large amplitude,
propagating droplets by numerically demonstrating that $J < 0$ and by
performing time-dependent numerical simulations of the LL Equation
(\ref{eq:landaulifshitz-vector}) with initially perturbed propagating
droplets.

\begin{figure}
  \centering
  \includegraphics[scale=1]{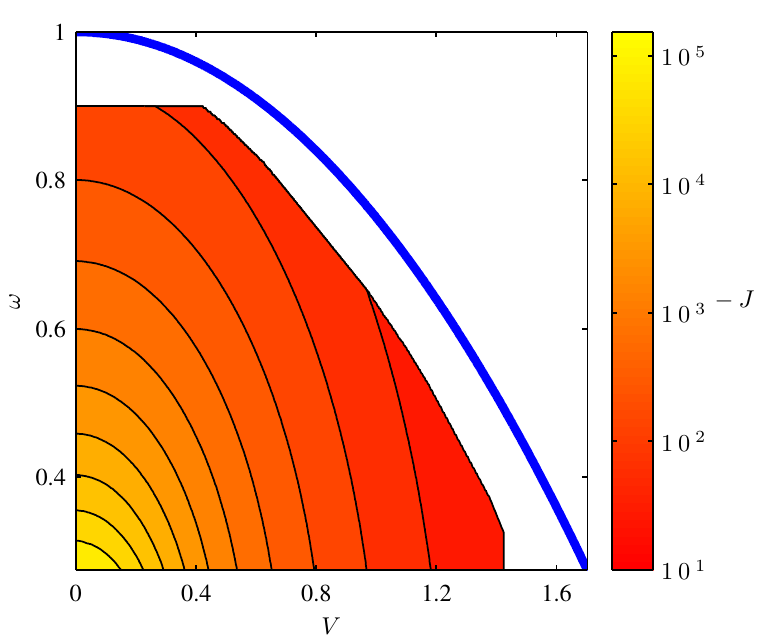}
  \caption{Numerical computation of the negative Jacobian $-J$ in
    Eq.~(\ref{eq:83}) demonstrating that the map $(\omega,V) \mapsto
    (\mathcal{N},\mathcal{P}_1)$ is invertible ($J \ne 0$) and the
    Jacobian condition $J < 0$ for stability is satisfied.}
  \label{fig:jacobian}
\end{figure}

Figure \ref{fig:jacobian} exhibits the numerical evaluation of the
negative Jacobian $-J$ (\ref{eq:83}) for $(V,\omega) \in S$.  Since $J
< 0$, we have numerically confirmed the invertibility of the map
$(\omega,V) \mapsto (\mathcal{N},\mathcal{P}_1)$ and the Jacobian
condition (\ref{eq:88}) for droplet stability.  The invertibility of
the map is not obvious given the lack of monotonicity of
$\mathcal{P}_1$ in Fig.~\ref{fig:contourplots-momentum}.  Note also
that the Jacobian is monotonic in both $\omega$ and $V$ within the set
$S$ and in the weakly nonlinear regime (Eq.~(\ref{eq:84})).

Recalling that the Jacobian condition $J < 0$ for orbital stability of
solitary waves has only been demonstrated in other nonlinear wave
systems
\cite{baryakhtar_vector_1983,ivanov_three-dimensional_1984,buryak_stability_1996},
we seek further evidence of stability by performing time-dependent
numerical simulations of perturbed droplets.  To solve
Eq.~(\ref{eq:landaulifshitz-vector}) numerically, we use the method of
lines with a pseudospectral, Fourier discretization in space, the same
as that used in our iterative numerical scheme of
Sec.~\ref{sec:numerical solution BVP}. For time stepping, we used an
adaptive time stepping, second order Runge-Kutta ODE solver.
Numerical parameters we used were $L = 50$, $N = 768$, giving $\Delta
\approx 0.13$.  Because the time stepping method does not preserve
the length of the magnetization vector, after each time step the
solution is rescaled so that $| \vec{m} | \equiv 1$ is preserved.  The
initial conditions consist of a computed droplet perturbed by noise
(explained below).  In all computations, we have verified that the
quantities (\ref{eq:3}), (\ref{eq:4}), and (\ref{eq:5}) change at most
by $0.74\%$ over the course of the simulation.

The initial conditions are constructed as follows.  Uncorrelated, mean
zero, normal random variables with variance 0.2 are sampled at each
grid point to construct a complex valued noise function $n$.  A low
pass filter is applied with cutoff wavenumber 6 yielding $\tilde{n}$.
Adding a droplet, $f$, computed as in Sec.~\ref{sec:numerical solution
  BVP} to the noise yields the perturbed droplet $\tilde{f} = f +
\tilde{n}$ in stereographic form.  Inverting the relation
(\ref{eq:eq:stereographicprojection})
\begin{equation*}
   m_3 = \frac{1 -
      |\tilde{f}|^2}{1 + |\tilde{f} |^2}\,, \quad m_1 + i m_2 = (1 +
    m_3) \tilde{f}\,,
\end{equation*}
yields the initial condition for the LL equation
(\ref{eq:landaulifshitz-vector}) in vector form.

\begin{figure}[!htb]
  \centering
  \includegraphics[scale=1]{
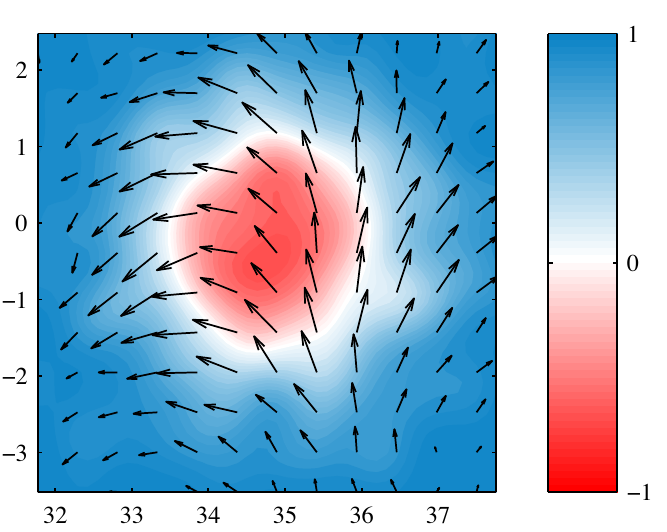}
  \caption{Perturbed droplet with $(V,\omega) = (0.7,0.3)$ at $t =
    50$.  Compare with unperturbed case in
    Fig.~\ref{fig:numericalsolution:bvp}.}
  \label{fig:perturbed_droplet}
\end{figure}

\begin{figure}[!htb]
  \centering \subfigure[{$x_1$ position of solution
    minimum.}\label{fig:noise_xmin}]%
  {{\includegraphics[scale=1]{
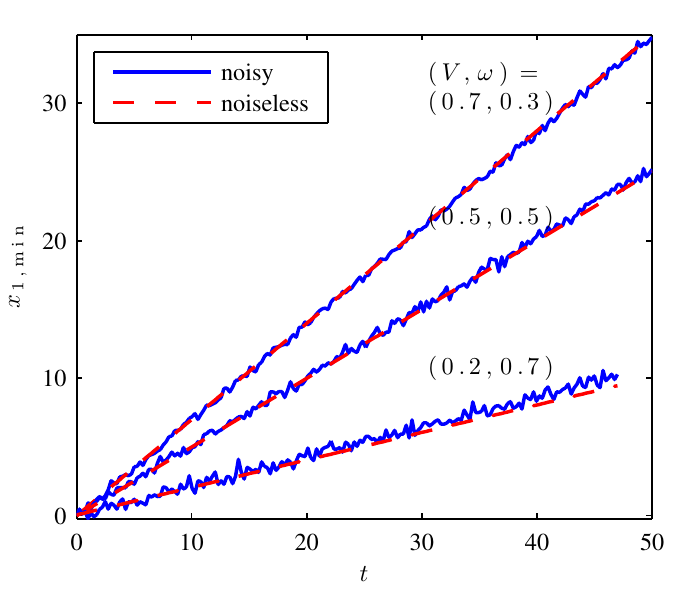}}}\\%
  \subfigure[{Phase at solution minimum.}\label{fig:noise_phase}]%
  {{\includegraphics[scale=1]{
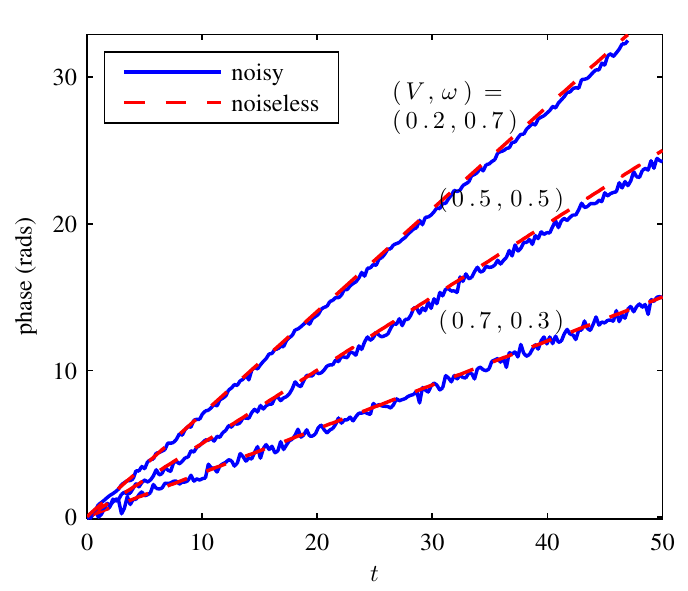}}}\\%
  \caption{Time evolution of the $x_1$ position (a) and the phase (b)
    at the minimum of three perturbed droplets (solid curves) compared
    with unperturbed droplet evolution (dashed
    lines).\label{fig:noise}}
\end{figure}

An example solution at $t = 50$ is shown in Fig.
\ref{fig:perturbed_droplet}.  The perturbed droplet's localization
structure has been maintained.  Furthermore, it's propagation speed
and precessional coherence have also been approximately maintained.
To demonstrate this, we show in Figs.~\ref{fig:noise_xmin} and
\ref{fig:noise_phase} the position and phase, respectively, of the
minimum of the computed solution as a function of time.  For
comparison, we have included the trajectories of these quantities for
unperturbed droplets.  Stability for the considered droplets is
clearly achieved with only small changes in the propagation speed and
rest frequency.
These results suggest that droplets are robust, approximately
maintaining their precessional frequency and propagation speed, even
after experiencing significant perturbations.

The droplet iterative scheme of Sec.~\ref{sec:numerical solution BVP}
converges for parameter values satisfying $\nu^2(V,\omega) < 0$.
However, as expected, such solutions are unstable due to linear spin
wave resonances.

\section{Discussion}
\label{sec:discussion}

In many respects, the 2D propagating nontopological droplet
considered here has much in common with the 1D propagating droplet
soliton reviewed in Sec. \ref{sec:1D droplet}. Moreover, by
rewriting Eqs.~(\ref{eq:23}) in polar coordinates, it is immediate
to observe that, for large values of the polar radius $\rho$, the
droplet behaves as a 1-dimensional object (namely, depending only on
$\rho$, with the velocity $V$ replaced by $V\,\cos{\varphi}$).

One interesting feature of the 1D case is the development of
topological structure, a phase singularity for $V = 0$ and $\omega <
0$.  While it is not possible for a single, topological solitary wave
to stably propagate \cite{papanicolaou_dynamics_1991}, the
copropagation of pairs with opposite charge are possible
\cite{papanicolaou_semitopological_1999}. The bifurcation of
nontopological droplet solitary waves to pairs of topological
structures have been predicted in isotropic ferromagnets
\cite{cooper_solitary_1998} and in 2D, uniaxial, easy-plane
ferromagnets \cite{papanicolaou_semitopological_1999}.  It is natural
to ask whether there exist propagating solutions with some local
topological structure for the LL equation
(\ref{eq:landaulifshitz-vector}).  Time-dependent numerical
simulations of an appropriately chosen initial value problem have
apparently yielded such structures \cite{piette_localized_1998}. Our
computations with the stereographic Eq.~(\ref{eq:f-equation}) are
limited to nonsingular, hence locally nontopological solutions.
Furthermore, the large momentum states we have computed exhibit large
phase gradients near the droplet center, suggesting the possibility of
a bifurcation into local topological structure for sufficiently small
or negative rest frequencies.

Another, necessarily multidimensional property of droplets was
studied in \cite{piette_localized_1998} where ninety degree scattering
was numerically observed for the head-on collision of two
``accelerated'' stationary droplets with the same frequency and
opposite velocities.  Merging accompanied by spin wave radiation was
also observed for certain initial configurations.  The propagating
droplets constructed in this work could be used as initial conditions
for a direct study of 2D droplet interactions.

\section{Conclusions\label{sec:conclusions}}%

Through numerical and asymptotic means, we have demonstrated the
existence and studied the properties of moving droplets across a wide
range of precessional frequencies and speeds as well as their
stability to perturbations.  Along with the rise of nanomagnetism
research and applications \cite{lau_magnetic_2011}, this work
encourages further study of localized magnetic structures including
droplet propagation in the presence of damping, external magnetic
fields, nonlocal dipolar fields, and injected spin current.

\section*{Appendix: Validation of the Numerical
  Scheme \label{sec:numerical validation}}%

We validate the numerical scheme described in Sec.
\ref{sec:numerical solution BVP} in several ways.  First, the
Derrick identity (\ref{eq:29}) was checked for the computed droplet
modes with $(V,\omega) \in S$.  The largest deviation from zero of
the quantity $d'(1)$ in (\ref{eq:29}) was found to be approximately
$5.3\cdot10^{-8}$, which occurred for the stationary droplet $V =
0$, $\omega = 0.9$, likely due to the finite domain size.

\begin{figure}
  \centering
  \includegraphics[scale=1]{
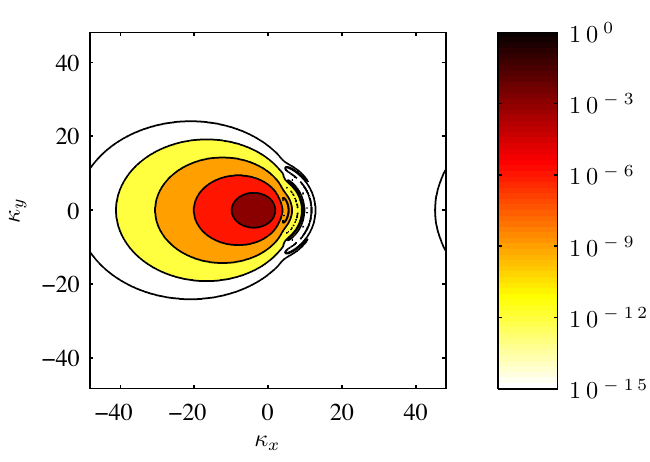}
  \caption{Absolute value of the Fourier coefficients,
    normalized by the maximum, for a single droplet with $\omega=0.3$
    and $V=0.212$. \label{fig:decay fourier coefficients}}
\end{figure}

Second, we verify that the resolution of the employed discretization
is sufficiently fine to resolve the solution's features. As can be
observed in Fig.~\ref{fig:decay fourier coefficients}, we confirm
that the relative magnitudes of the Fourier coefficients decay to
machine precision for a large momentum droplet solution with
$V=0.212$ and $\omega=0.3$.  We find that all droplets with
$(V,\omega) \in S$ show similar behavior with a large momentum
droplet exhibiting the slowest decay to at most $4.9 \cdot 10^{-9}$
for large $\kappa$, still an acceptable value.  This suggests
negligible aliasing effects so that we have properly resolved the
solution.

\begin{figure}
  \centering
  \includegraphics[scale=1]{
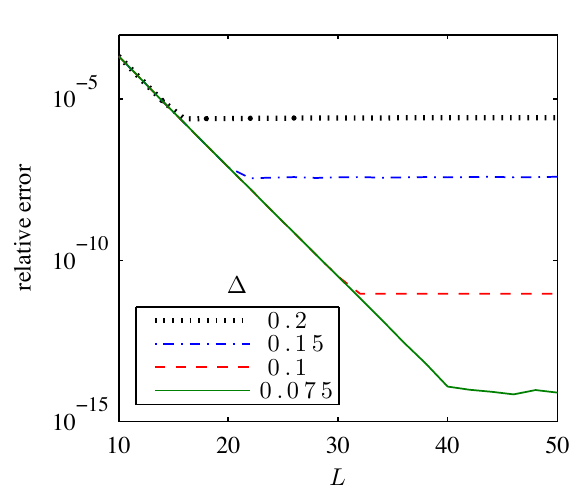}
  \caption{Relative error of the solution of the boundary value
    problem as a function of the domain size ($L_1=L_2=L$) for
    different values of the grid spacing ($\Delta_1 = \Delta_2
    =\Delta$) for a single droplet with $\omega=0.3$ and
    $V=0.7$. \label{fig:droplet accuracy}}
\end{figure}

Finally, to determine a domain size and grid resolution that
accurately resolves the droplets, we computed a very accurate,
yardstick solution, $f_{\mathrm{ys}}$, with $V=0.7$ and $\omega=0.3$.
The numerical parameters used were $L=50$, $N=2048$, $c=10$, and
$\mathrm{tol}_{f}=\mathrm{tol}_r=\mathrm{tol}_{\eta}=10^{-13}$ so that
$\Delta \approx 0.0488$.  We compare $f_{\mathrm{ys}}$ with the same
droplet mode obtained on different grids $(\Delta,L)$.
The relative error $\|f_{\mathrm{ys}}-f \|_\infty/
\|f_{\mathrm{ys}}\|_\infty$ is computed using sinc interpolation
\cite{frank_summary_2000} of the yardstick to the grid for $f$ and
plotted in Fig.~\ref{fig:droplet accuracy}.  Domain truncation error
decreases with increasing domain size until discretization error
dominates.  The smallest grid spacing $\Delta = 0.075$ with $L > 40$
yields relative errors below the precision of the yardstick
tolerances.  The iterative computations of droplets with $(V,\omega)
\in S$ employ the conservative values $(\Delta,L) = (0.065,50)$,
accounting for the larger phase gradients that occur.



\end{document}